\documentstyle[12pt]{article} 
\newcommand{\be}{\begin{equation}}
\newcommand{\ee}{\end{equation}}
\newcommand{\ba}{\begin{array}}
\newcommand{\ea}{\end{array}}
\newcommand{\bea}{\begin{eqnarray}}
\newcommand{\eea}{\end{eqnarray}}
\newcommand{\beaa}{\begin{eqnarray*}}
\newcommand{\eeaa}{\end{eqnarray*}}
\newcommand{\al}{\alpha}

\newcommand{\de}{\delta}
\newcommand{\De}{\Delta}
\newcommand{\ep}{\epsilon}

\newcommand{\Ga}{\Gamma}
\newcommand{\la}{\lambda}

\newcommand{\si}{\sigma}
\newcommand{\th}{\theta}

\newcommand{\rb}{\right]}
\newcommand{\lb}{\left[}
\newcommand{\lag}{\langle}
\newcommand{\rag}{\rangle}
\newcommand{\bl}{\biggl(}
\newcommand{\br}{\biggr)}
\newcommand{\cbl}{\biggl\{ }
\newcommand{\cbr}{\biggr\} }
\renewcommand{\(}{\left(}
\renewcommand{\)}{\right)}
\newcommand{\ao}{a_n^{(1)}}
\newcommand{\ato}{a_2^{(1)}}
\newcommand{\atn}{a_{2n}^{(2)}}

\newcommand{\bo}{b_n^{(1)}}

\newcommand{\dt}{d_{n+1}^{(2)}}

\newcommand{\uq}[1]{$U_q(#1)$}

\newcommand{\cR}{\check{R}}
\newcommand{\cP}{\check{P}}

\newcommand{\tbe}{\tilde{\beta}}

\newcommand{\nn}{\nonumber}
\newcommand{\fns}{\footnotesize}
\newcommand{\scs}{\scriptsize}
\newcommand{\shs}{\shortstack}
\newcommand{\ol}{\overline}
\newcommand{\ul}{\underline}
\newcommand{\noi}{\noindent}
\newcommand{\hs}{\hspace}
\newcommand{\vs}{\vspace}
\newcommand{\lra}{\longrightarrow}
\newcommand{\ot}{\otimes}
\newcommand{\tA}{\tilde{A}}
\newcommand{\tB}{\tilde{B}}
\newcommand{\tC}{\tilde{C}}
\newcommand{\cA}{{\cal A}}
\newcommand{\cB}{{\cal B}}
\newcommand{\cC}{{\cal C}}
\newcommand{\cD}{{\cal D}}
\newcommand{\tcA}{\tilde{\cal A}}
\newcommand{\tcB}{\tilde{\cal B}}
\advance\textheight by \topskip
\renewcommand{\baselinestretch}{1.3} 

\makeatletter

\@addtoreset{equation}{section}
\def\section{\@startsection {section}{1}{\z@}{-8.5ex plus -1ex minus
 -.2ex}{3.3ex plus .2ex}{\large\bf\centering}}
\def\subsection{\@startsection{subsection}{2}{\z@}{-3.25ex plus
 -1ex minus -.2ex}{1.5ex plus .2ex}{\bf}}
\def\subsubsection{\@startsection{subsubsection}{3}{\z@}{-3.25ex plus%
 -1ex minus -.2ex}{1.5ex plus .2ex}{\sl}}
\begin{document}

\newpage
\begin{titlepage}
\begin{flushright}
BRX-TH-408\\
DAMTP/97-25\\
hep-th/9703158\\
{\bf March 1997}
\end{flushright}
\vspace{1cm}
\begin{center}
{\Large {\bf 
Trigonometric S-matrices, Affine Toda Solitons
\vspace{.2cm}\\
and Supersymmetry
}}\\
\vspace{1.5cm}
{\large Georg M.\ Gandenberger}\footnote{\noi E-mail:
gandenberger@binah.cc.brandeis.edu \\  On leave from:
Department of Mathematical Sciences, Durham University, Durham DH1 3LE, U.K.\\  
(e-mail: G.M.Gandenberger@durham.ac.uk)}
\vspace{3mm}
\vs{0.5cm}

{\em Department of Physics, Brandeis University,}\\
{\em Waltham, MA - 02254, USA}\\
\vspace{2cm} {\bf{ABSTRACT}}
\end{center}
\begin{quote}
Using $U_q(a_n^{(1)})$-- and $U_q(a_{2n}^{(2)})$--invariant
$R$-matrices we construct exact \mbox{$S$-matrices} in
two--dimensional space--time. 
These are conjectured to describe the scattering of solitons in affine
Toda field theories. In order to find the spectrum of soliton bound
states  we examine the pole structure of these
$S$-matrices in detail. We also construct the $S$-matrices for all
scattering processes involving scalar bound states. In the last part
of this paper we discuss the connection of these $S$-matrices with
minimal $N=1$ and $N=2$ supersymmetric $S$-matrices. In particular we
comment on the folding from $N=2$ to $N=1$ theories.   

\end{quote}

\vfill

\end{titlepage}

\newpage

%
%
\section{Introduction}

During the last twenty years, a great deal of effort has been expended
on
the study of integrable two--dimensional field theories. Unlike their
higher dimensional counterparts, these theories can often be solved
exactly without the need to resort to perturbative methods. It is for
this reason that two--dimensional field theories can provide important
insights which may lead towards a better understanding of
non--perturbative aspects of general quantum field theories. Apart
{}from their role as models for higher dimensional theories,
two--dimensional theories also have some 
important applications in string theory. In particular, the study of
supersymmetric (and superconformal) theories is closely related to the
study of world sheet supersymmetries in superstring theories.  

This paper deals with the construction of exact $S$-matrices for
a certain class of relativistic quantum field theories defined 
in $(1+1)$-dimensional space--time, namely theories displaying a
quantum affine symmetry. We also briefly discuss the general structure
of $S$-matrices for supersymmetric quantum field theories and point
out some interesting connections between the trigonometric $S$-matrices
and supersymmetric $S$-matrices.  The construction of the
$S$-matrices in this paper follows closely the construction in the
series of papers \cite{hollo93,gande95,gande95b,gande96}. A more
detailed exposition of some of the material presented here can also be
found in \cite{gande96c}. 

The layout of this paper is as follows. In the first
chapter we provide a brief introduction to some of the main results of
the study of affine Toda field theories (ATFTs). A 
review of the subject of trigonometric $R$-matrices and exact
\mbox{$S$-matrices} for theories with 
quantum affine symmetries is given in chapter 2. Chapter 3 deals with
the construction of exact $S$-matrices using \uq{\ao} invariant
$R$-matrices. We examine 
the pole structure of these $S$-matrices and determine scattering
amplitudes for the scattering of bound states. We also summarise the
evidence for the conjecture that these $S$-matrices describe the
scattering of solitons in $\ao$ ATFTs. In \mbox{chapter 4} we repeat
this construction for the case of \uq{\atn} invariant $R$-matrices and
$\atn$ affine Toda solitons.
Chapter 5 deals with a slightly separate subject, namely the minimal
$S$-matrices for two-dimensional $N=1$ and $N=2$ supersymmetric
models. We find, however that the \mbox{$S$-matrix} scalar factors in these
supersymmetric cases are related to those in chapters 3 and 4 at
one particular value of the coupling constant. We also comment briefly
on a possible ``folding'' from $N=2$ to $N=1$ theories as discussed
recently by Moriconi and Schoutens \cite{moric95}.

\subsection{Affine Toda field theories}

Affine Toda field theories (ATFTs) are a family of (classically) integrable
field theories in $(1+1)$--dimensional Minkowski space. They are
defined by their Lagrangian density
\be
{\cal L} =
\frac12(\partial_{\mu}\phi^a)(\partial^{\mu}\phi^a) -
\frac{m^2}{\tilde{\beta}^2} 
\sum_{j=0}^r n_je^{\tilde{\beta}\al_j\phi}\;, \label{lagrangian}
\ee
in which $\phi = (\phi^1,...,\phi^r)$ is a $r$-dimensional scalar field,
the $\al_i$'s are a set of $r+1$ \mbox{$r$-dimensional} vectors and
$m$ and $\tilde{\beta}$ are mass and coupling constant parameters.
\{$\al_1,...,\al_r$\} form the root system of a semi--simple classical
Lie algebra $g$ of rank $r$, and $\al_0$ is chosen to be the extended
root, such that \{$\al_0,\al_1,\dots,\al_r$\} form the root system of
the affine Lie algebra $\hat{g}$ (the longest root is taken to have
length $\sqrt{2}$, except in the case of a twisted Lie algebra
$\hat{g}^{(k)}$ in which we take the longest root to have length
$\sqrt{2k}\;$). The numbers $n_j$ are the Kac marks of the affine
algebra $\hat{g}$ and we have chosen $n_0 =1$. 
Thus, there is one ATFT associated with each simple affine Lie algebra
$\hat{g}$.  
Among the ATFTs one has to distinguish between
the two fundamentally different cases, the so--called real and
imaginary ATFTs, which are distinguished by the coupling constant
$\tbe$ being either a real or purely imaginary number.

In the case of a real coupling constant $\tbe$ the Lagrangian
describes a unitary field theory containing $r$ scalar
particles. The classical masses and three point
couplings for all real ATFTs were computed in \cite{brade90} and it
was found that the particle masses form the right Perron--Frobenius
eigenvector of the Cartan matrix of the underlying algebra. This fact
suggested that each particle can be associated with one of the nodes
of the Dynkin diagram.

It is generally believed that the integrability of these theories is
preserved after quantisation and that ATFTs possess exact factorised
$S$-matrices. The masses of the particles in real ATFTs are
non--degenerate and the $S$-matrices are therefore diagonal. By
using the axioms of analyticity, unitarity, crossing symmetry and the
bootstrap, exact $S$-matrices for the simply--laced ATFTs (i.e.\ based
on the $a,d$ and $e$ series of affine Lie algebras) have been
constructed by Braden et al.\ in \cite{brade90} and independently by
Christe and Mussardo in \cite{chris90}. The
\mbox{$S$-matrices} for the non--simply--laced and twisted algebras,
in which the poles corresponding to bound states require an additional
coupling constant dependence, were found some time later by Delius et
al.\ in \cite{deliu92}. 
One of the most intriguing features of these $S$-matrices is the fact
that they exhibit a strong--weak coupling duality. This means that if
we change the coupling constant in the $S$-matrices for the $\hat{g}$
ATFT in the following way 
\be
\tbe \mapsto \frac{4\pi}{\tbe}\;,
\ee
then we recover the $S$-matrices of the ATFT associated with the dual
algebra $\hat{g}^{\vee}$. (The dual of an affine Lie algebra is
obtained by exchanging long and short roots.)

\subsection{Affine Toda solitons}

Let us now take the coupling constant in (\ref{lagrangian}) to be
purely imaginary, i.e.\ $\tbe = i\beta$ with $\beta$ being a real
number. 
This seemingly small change in the Lagrangian has some very
significant implications for these theories. 
First of all, apart from the $a_1^{(1)}$ theory, which is the well
known Sine--Gordon theory, the imaginary ATFTs are in general
non--unitary.

The most striking feature of imaginary ATFTs, however, is the fact
that they display a infinitely degenerate vacuum and therefore admit
classical soliton solutions. By using Hirota's method 
for the solution of nonlinear differential equations, Hollowood was
able in \cite{hollo92} to construct explicit expressions for
one--soliton and multi--soliton solutions in $a_n^{(1)}$ ATFTs.
Some time later Olive et al.\ \cite{olive93}
used a more algebraic construction in order to obtain the general
solution to the classical equations of motion for ATFTs based on all
affine algebras. Despite the complex form of the Hamiltonian, the
solitons turn out to have real and positive energies and masses.  
As pointed out by Olive et al., this fact suggests that there might be
a unitary theory embedded in the (generally non--unitary) imaginary
ATFTs. Remarkably, the masses of the solitons were also found to be
proportional to the particle masses in the dual theory\footnote{Note
   that by dual theory we mean in this case the
   theory obtained after the exchange of the underlying simple Lie
   algebra $g$ with its dual $g^{\vee}$, whereas the weak--strong
   coupling duality of the $S$-matrices relates the affine algebra
   $\hat{g}$ with its affine dual $\hat{g}^{\vee}$.}. 
 
Apart from the elementary solitons, there are also bound states of
solitons which can be regarded as two solitons oscillating around a
fixed point. Whereas in Sine--Gordon theory only bound states with zero
topological charge (the `breathers') occur, in most other ATFTs there
are also bound states with non--zero topological charges (the
`breathing solitons' or `excited solitons'). In \cite{harde95} Harder
et al.\ obtained classical bound state solutions for $a_n^{(1)}$ ATFTs
by changing the real velocity into an imaginary velocity in the
expressions of the two--soliton solutions. They found scalar bound
states, which are the analogues to the Sine--Gordon breathers, and also
bound states transforming under the $2a$th (for $a=1,2,\dots
<(n+1)/2$) and $(2a-n-1)$th (for $(n+1)/2 < a < n$) fundamental
representations of $\ao$. While in the classical theory a continuous
spectrum of bound states exists, in the quantum theory we expect the
spectrum to be quantised and a finite number of discrete bound states
to emerge. 

Unlike in the case of real coupling constant the quantum theory of the
imaginary ATFTs is not very well established. This is mainly due to
the fact that the Lagrangian in this case defines, in general, a
non--unitary field theory. Despite this problem, attempts have been
made to construct exact $S$-matrices for the scattering of quantum
solitons in ATFTs. These attempts rest primarily on the assumption of
a quantum affine symmetry, which has been established in
\cite{berna90,felde92} for certain restrictions of imaginary ATFTs. As
we will review in the following chapter such a quantum affine symmetry
permits the use of
trigonometric $R$-matrices as scattering matrices. Hollowood was the
first to achieve the explicit construction of exact $S$-matrices for
$\ao$ affine Toda solitons by using \uq{\ao} invariant
$R$-matrices. Following the realisation of the crucial importance of 
$R$-matrix gradations in \cite{brack94,deliu95b} this construction was
extended to the cases of $\dt$ and $\bo$ ATFTs in
\cite{gande95b,gande96}. In two very recent papers \cite{takac97} the
same construction was used for the two exceptional cases of
$d_4^{(3)}$ and $g_2^{(1)}$ ATFTs. 
In the present paper we extend Hollowood's construction for the
$a_n^{(1)}$ case to the scattering of bound states of solitons and we
construct the $S$-matrices for the case of the twisted algebra
$\atn$.

%
%
\section{Exact $S$-matrices from trigonometric $R$-matrices}

In this section we provide a short introduction to the subject of
$R$-matrices of quantised universal enveloping algebras (QUEAs) of
affine Lie algebras. These $R$-matrices are
trigonometric solutions of the quantum Yang--Baxter equation. 
Although they were not originally introduced for this purpose,
trigonometric $R$-matrices turn out to be exactly the right objects 
to use as $S$-matrices for theories displaying a so--called quantum
affine symmetry.   
Following Delius \cite{deliu95b} (see that paper for details) we give
a general scheme of how to build exact soliton $S$-matrices in terms
of these trigonometric \mbox{$R$-matrices}.  

\subsection{$R$-matrices and the Yang Baxter Equation}

Associated with each QUEA $U_q(\hat g)$ there exists a universal
$R$-matrix ${\cal R} \in U_q(\hat g)\ot U_q(\hat g)$, which satisfies
the following identities:
\bea
(\De\ot I){\cal R} = {\cal R}_{13}{\cal R}_{23}\;,&&\;\;\;\;  
(I\ot\De){\cal R} = {\cal R}_{13}{\cal R}_{12}\;,
\label{hopfaxiom1} \\
\ol{\De}(a){\cal R} = {\cal R}\De(a)&& \hs{0.6cm} (\forall a\in
U_q(\hat g) )\;,
\label{hopfaxiom2} \eea 
in which $\De$ is the (Hopf algebra) coproduct $\De: U_q(\hat g) \to
U_q(\hat g) \ot U_q(\hat g)$, $\ol{\De}$ is the opposite coproduct
defined by $\ol{\De} = T\circ\De$ and $T$ denotes 
the transposition map, i.e.\ $T(a_1\ot a_2) = a_2\ot a_1$
($\forall a_1,a_2 \in U_q(\hat g)$).  If we write 
${\cal R} = \sum_n {\cal R}_n^{(1)}\ot {\cal R}_n^{(2)}$, then 
\mbox{${\cal R}_{ij} \equiv \sum_n 1\ot \dots\ot{\cal
R}_n^{(1)}\ot 1\ot \dots\ot {\cal R}_n^{(2)}\ot\dots\ot 1$}, 
in which ${\cal R}_n^{(1)}$ and ${\cal 
R}_n^{(2)}$ appear in the $i$th and $j$th position. 
It is also possible to deduce the following identities, which later
will be used in connection with the crossing symmetry of $S$-matrices: 
\be 
(\ep\ot I){\cal R} = (I \ot \ep){\cal R} = 1
\ee
and
\be
(S\ot I){\cal R} = {\cal R}^{-1}\;,\;\;\;\;
(I\ot S){\cal R}^{-1} = {\cal R}\;,\label{rmcros}
\ee
in which $I$, $\ep$ and $S$ denote the identity, counit and antipode
in $U_q(\hat g)$, respectively. For our purpose
the most important property of universal
$R$-matrices is the fact that they satisfy the quantum
Yang-Baxter equation (YBE) 
\be
{\cal R}_{12}{\cal R}_{13}{\cal R}_{23} = {\cal R}_{23}{\cal
R}_{13}{\cal R}_{12}\;. \label{aqybe}
\ee
Instead of the universal $R$-matrices, here we require expressions of
$R$-matrices which depend on a spectral parameter and on some finite
dimensional representations of $U_q(\hat g)$.  
In order to achieve this we first define an automorphism $D_x^s: U_q(\hat
g) \to U_q(\hat g)$ which acts on the Chevalley generators $e_i,f_i$
and $h_i$ of \uq{\hat g} in the following way:  
\be
D_x^s(e_i) = x^{s_i}e_i, \hs{20pt} D^s_x(f_i) =
x^{-s_i}f_i, \hs{20pt} D^s_x(h_i) = h_i, \label{dtrafo}
\ee 
in which $x$ is a complex number and the $s_i$ (i=0,1,\dots,n) are a
set of real numbers which determine the gradation $s$ of the
$R$-matrix. The most commonly used gradations are the homogeneous
gradation, in which $s_i=\de_{i0}$, and the principal gradation, in
which $s_i=1\;, \hs{0.2cm} (\forall i=0,1,\dots,n)$.
Thus we can define a $R$-matrix depending on a spectral
parameter $x$ and on the gradation $s$:
\be 
{\cal R}^{(s)}(x) \equiv (D_x^s \ot I){\cal R}\;.\label{repindrm}
\ee
Now let $V_i$ and $V_j$ be two finite dimensional $U_q(\hat
g)$-modules and \mbox{$\pi_{i(j)}: U_q(\hat g) \to
\mbox{End}(V_{i(j)})$} the associated representations, and we define
\be
\cR^{(s)}_{i,j}(x) \equiv \si_{ij}(\pi_i \ot \pi_j){\cal R}^{(s)}(x),
\label{rmatrix} 
\ee
in which $\si_{ij}$ is a permutation operator $\si_{ij}(v\ot w) =
w\ot v \hs{5pt}  (\forall v\in V_i, w \in V_j)$. Thus
$\cR^{(s)}_{i,j}(x)$ acts as an intertwiner on the spaces $V_i$ and
$V_j$: 
\be 
\cR^{(s)}_{i,j}(x): V_i\ot V_j \to V_j\ot V_i\;.
\label{intertwiner}
\ee 
Note that $R$-matrices in different gradations can be related to
each other through a gauge transformation, i.e.\ given two gradations
$s^1$ and $s^2$ we can always find a transformation
$\tau_{ij}:V_i\ot V_j \lra V_i\ot V_j$, such that
$\cR^{(s^1)}_{i,j}(x) = \tau_{ji}\cR^{(s^2)}_{i,j}(x)\tau_{ij}^{-1}$. 
In the remainder of this paper we will mainly use $R$-matrices
in the homogeneous gradation $s^h$. For the sake of simplicity we
therefore omit the gradation label on $R$-matrices in the homogeneous
gradation, i.e.\ $\cR_{i,j}(x) \equiv \cR^{(s^h)}_{i,j}(x)$.  For any
three finite dimensional representations the associated YBE now takes
the form 
\bea
\lefteqn{\lb \cR_{j,k}(x)\ot I_i\rb \lb I_j\ot
\cR_{i,k}(xy)\rb \lb \cR_{i,j}(y)\ot I_k\rb =}
\hs{3.5cm} \nn \\
& & = \lb I_k\ot \cR_{i,j}(y)\rb \lb \cR_{i,k}(xy)\ot
I_j \rb \lb I_i\ot \cR_{j,k}(x)\rb \;, \label{ybe}
\eea
in which $I_l$ denotes the identity on $V_l$ and therefore both sides
of equation (\ref{ybe}) map $V_i\ot V_j\ot V_k$ into
$V_k\ot V_j\ot V_i$.  Furthermore, let us assume that we can write the
tensor product of two modules $V_i$ and $V_j$ as a direct
sum of irreducible modules in the following way
\be
V_i\ot V_j = \bigoplus_k V_k\;. \label{tensprod}
\ee 
It can then be shown that the general solution to equation (\ref{ybe})
can be written as a spectral decomposition
\be
\cR_{i,j}(x) = \sum_k \rho_k(x) \cP^{ij}_k, \label{rspecdecom}
\ee
in which $\cP^{ij}_k: V_i\ot V_j \to V_k \subset V_j\ot V_i$ denotes
the intertwining projector onto the irreducible module $V_k$ and the
$\rho_k(x)$ are scalar factors. Accordingly, it is also immediately
evident that these $R$-matrices
satisfy the following normalisation condition:
\be 
\cR_{a,b}(x)\cR_{b,a}(x^{-1}) = I_b \ot I_a \;. \label{rmnormalisation}
\ee
Using a method known as the tensor product graph method it is possible
to find explicit expressions of trigonometric $R$-matrices in the form
(\ref{rspecdecom}). At this stage, however, this method is only
applicable to the case of multiplicity--free tensor products, i.e.\
those cases in which each $V_k$ appears only once at the right hand
side of (\ref{tensprod}). 

\subsection{Exact $S$-matrices with quantum affine symmetries}

Let us assume that we have a relativistic quantum field theory in
$(1+1)$-dimensional space--time. Following Delius \cite{deliu95b}, we
say this theory has a \uq{{\hat g}} {\bf quantum affine symmetry} if the
following two conditions hold: 
\begin{quote}
\noi 1) The theory has quantum conserved charges $H_i$, $E_i$, $F_i$,
for $(i=1,2,...,n)$, which satisfy the same commutation relations as
the Chevalley generators of \uq{\hat g}.  

\noi 2) These conserved charges possess definite Lorentz spin.
If we denote the infinitesimal Lorentz generator by $D$, the
conserved charges transform under $D$ in the following way
\be
D(E_i) = s_i E_i\;,\;\;\; D(F_i) = -s_i F_i\;,\;\;\; D(H_i) = 0\;,
\label{inflorentz} 
\ee
in which $s_i$ and $-s_i$ are the Lorentz spins of the conserved charges
$E_i$ and $F_i$ respectively. 
\end{quote}

\noi In other words, we can find conserved charges which act as the
generators of an \uq{\hat g} charge algebra, in which ${\hat g}$ is a
rank $n$ affine Lie algebra. And one finds that the Lorentz spin of
these conserved charges plays the role of the derivation in $U_q(\hat
g)$. Thus we obtain that the gradation of the charge algebra is
determined by the  Lorentz spins $s_i$ of the conserved charges. We
will therefore call this gradation the {\bf spin gradation}. 

As Delius pointed out in \cite{deliu95b}, the existence of a quantum
affine symmetry also implies the complete integrability of the theory.
Thus we assume that we are dealing with an integrable theory which
possesses factorised $S$-matrices, which means that all the
information of any multi--particle scattering process is contained in
the two-particle $S$-matrices. Let us further assume that we can
arrange the quantum states in the theory into mass degenerate
multiplets which are the modules of some finite dimensional
irreducible representations of \uq{\hat g}. Let us denote these
multiplets by $V_1,V_2,\dots,V_n$. Thus the two--particle $S$-matrices
must act as intertwiners on these representation spaces    
\be
S_{a,b}(\th): V_a\ot V_b \lra V_b\ot V_a\;,
\ee
and depend only on the rapidity difference $\th$ of the incoming
particles. These $S$-matrices are severely constrained by the 
so--called axioms of analytic $S$-matrix theory, which we briefly
review in the following:
\vs{4pt}

\noi {\em (i) Factorisation equation}:
\bea
\lefteqn{\lb S_{b,c}(\th_1)\otimes I_a\rb \lb I_b\otimes
S_{a,c}(\th_1+\th_2)\rb \lb S_{a,b}(\th_2)\otimes I_c\rb =}
\hspace{2cm} \nn \\
& & = \lb I_c\otimes S_{a,b}(\th_2)\rb \lb S_{a,c}(\th_1+\th_2)\otimes 
I_b \rb \lb I_a\otimes S_{b,c}(\th_1)\rb \;, \label{factorisation}
\eea
in which $\th_1 \equiv \th_b-\th_c$ and
$\th_2 \equiv \th_a-\th_b$. The factorisation equation is a result
of the existence of higher spin conserved charges in an integrable
theory.    
\vs{4pt}

\noi {\em (ii)  Unitarity}: \vs{2pt} \\
The unitary condition can be expressed as
\be
S_{a,b}(\th)S_{b,a}(-\th) = I_b \otimes I_a\;. \label{unitarity}
\ee

\noi {\em (iii) Crossing Symmetry}:\vs{2pt} \\
The matrix elements $S_{a,b}(\th)$ must be symmetric under the
transformation $\th \mapsto i\pi -\th$, such that
\bea
S_{a,b}(\th) &=& (I_b \otimes C_{\ol a})[\si_{\ol ab} S_{b,\ol
a}(i\pi-\th)]^{t_2} \si_{\ol ab} (C_a \otimes I_b) \nn \\
&=& (C_{\ol b} \otimes I_a)[\si_{a\ol b} S_{\ol b,a}(i\pi-\th)]^{t_1}
\si_{a\ol b} (I_a \otimes C_b)\;, \label{crossing}
\eea
in which $C_a:V_a \to V_{\ol a}$ is the charge conjugation operator
which maps the particles in $V_a$ to their charge conjugated partners
in $V_{\ol a}$ and $t_1$ ($t_2$) means transposition in the first
(second) space. 
\vs{4pt} 
  
\noi {\em (vi) Analyticity and Bootstrap}:\vs{2pt} \\
$S_{a,b}(\th)$ is a meromorphic function of $\th$ with the only
singularities on the physical strip \mbox{$(0\leq \rm{Im}\th \leq
\pi)$} at $\rm{Re}\th = 0$. Simple poles in the physical strip
correspond to bound states in the direct or crossed channel.
If $S_{a,b}(\th)$ exhibits a simple pole $\th_{ab}^c$ on the physical
strip corresponding to a bound state in $V_c$ ($\subset V_a\ot V_b\,$)
in the direct channel, then the mass of this particle in $V_c$ is
given by the formula: 
\be 
m_c^2 = m_a^2 + m_b^2 + 2m_am_b\cos(\mbox{Im}\th_{ab}^c)\;. \label{mass}
\ee
In this case there must also be
poles  $\th_{\ol ca}^{\ol b}$ in $S_{\ol c,a}(\th)$ and $\th_{b\ol
c}^{\ol a}$  in $S_{b,\ol c}(\th)$, such that $\th_{ab}^c + \th_{\ol
ca}^{\ol b} + \th_{b\ol c}^{\ol a} = 2\pi i$.
The bootstrap equations express the fact that there is no
difference whether the scattering process with any particle in, say,
$V_d$ occurs before or after the fusion of particles  $a$ and $b$ into
particle $c$:  
\be
S_{d,c}(\th) = [I_b \otimes S_{d,a}(\th-(i\pi-\th_{\ol ca}^{\ol b}))]
[S_{d,b}(\th+(i\pi-\th_{b\ol c}^{\ol a})) \otimes I_a]\;,
\label{bootstrap}  
\ee
in which both sides are restricted to $V_d \ot V_c \subset V_d \ot V_a
\ot V_b$.
\vs{6pt}

If we now compare these axioms with the properties of the $R$-matrices
as defined in (\ref{rmatrix}), we find that the $R$-matrices, which
are also intertwining maps on the representation spaces, already
satisfy most of the $S$-matrix axioms.  
The most important of these axioms is the factorisation equation,
which is the same as the YBE with additive spectral parameter
$\th$. However, the \mbox{$R$-matrix} satisfies the YBE (\ref{ybe})
with multiplicative spectral parameter $x$.  This difference is easily
circumvented if we choose $x(\th) = \exp(\th)$. 
The $R$-matrix normalisation (\ref{rmnormalisation}) now becomes
\be
I_b\ot I_a = \cR_{a,b}(x)\cR_{b,a}(x^{-1}) =
\cR_{a,b}(x(\th))\cR_{b,a}(x(-\th))\;,   
\ee
which is identical to the unitarity condition (\ref{unitarity}) of the
$S$-matrix. Furthermore, the equations (\ref{hopfaxiom1}) for the
$R$-matrices appear to be analogous to the bootstrap equations
(\ref{bootstrap}). The only condition which is not directly satisfied by
the $R$-matrices is the crossing symmetry
condition. However, it was also demonstrated in \cite{deliu95b} that the
crossing symmetry condition can be satisfied by choosing an overall
scalar factor (although care must be taken that this scalar factor
does not violate the unitarity condition). This `crossing property' of
the $R$-matrices is essentially implied by equations
(\ref{rmcros}).    

We therefore make the following general ansatz for the two-particle
$S$-matrix of a relativistic field theory displaying an \uq{\hat g}
quantum affine symmetry: 
\be
S_{a,b}(\th)\; =\; {\cal F}_{a,b}(\th)\; \cR^{(s)}_{a,b}(x)\;,
\label{genSM} 
\ee
in which ${\cal F}_{a,b}(\th)$ is an overall scalar factor,
$\cR^{(s)}_{a,b}(x)$ is the \uq{\hat g} invariant $R$-matrix in the spin
gradation, and the spectral parameter is given by $x= e^{\th}$. The
scalar factor ${\cal F}_{a,b}$ is determined by the requirements of
crossing symmetry and unitarity.  

In order to use this general ansatz for specific models we need to 
find an expression for the gauge transformation which transforms the
homogeneous gradation $R$-matrices into $R$-matrices in the `physical'
spin gradation. This is necessary since we only know exact expressions
of the \uq{\hat g} invariant $R$-matrices in the homogeneous gradation.
This has first been done by Bracken et al.\ in \cite{brack94}, in
which the following formula was found 
\be
{\cal R}^{(s)}(x) = \(x^{\nu\xi}\ot I\) {\cal R}^{(h)}(x^{\nu})
\(x^{\nu\xi}\ot I\)^{-1}\;, 
\ee
in which $\xi$ is some linear combination of the conserved
charges $H_1,H_2,\dots,H_n$. In particular it was found that the
change of gradation shifts the spectral parameter $x$ to $x^{\nu}$,
and $\nu$ is given by $\nu = \sum_{i=0}^n n_i s_i$, in which the
$n_i$'s are the Kac marks of the affine symmetry algebra and the
$s_i$'s are the Lorentz spins of the conserved charges.    
Hence, if we write the spectral parameter in the spin gradation
$R$-matrix as $x=e^{\th}$ we obtain the spectral parameter of the
homogeneous gradation $R$-matrix as
\be
x=\exp\(\sum_{i=0}^n n_i s_i \th \)\;. \label{xth1}
\ee  
(For details of this derivation see \cite{brack94,deliu95b} or
\cite{gande96c}).    

The only remaining problem concerns the role of the deformation
parameter $q$ in \uq{\hat g}. Heretofore $q$ has been an arbitrary
complex number (although not a root of unity). Clearly, if we choose
$R$-matrices as $S$-matrices the deformation parameter $q$, like the
spectral parameter $x$, must somehow be determined by physical quantities. 
In the example of  ATFTs in the following section we will see that $q$
is related to the coupling constant in the theory.

\subsection{$S$-matrices for affine Toda solitons}\label{ssecsmat}  

The question of whether there is a quantum affine symmetry present in
ATFTs with imaginary coupling constant has not
yet been answered satisfactorily. However, some strong evidence exists
suggesting that an ATFT with imaginary coupling
constant based on a rank $n$ affine Lie algebra ${\hat g}$ displays a 
\uq{{\hat g}^{\vee}} quantum affine symmetry, in which ${\hat
g}^{\vee}$ denotes the dual algebra of $\hat g$.    
This evidence emerges from the work by Bernard and LeClair
\cite{berna90} which was later followed up by  Felder and LeClair
\cite{felde92}. 
These papers discuss restrictions of imaginary ATFTs which
can be regarded as certain integrable perturbations of conformal field
theories.  This allows the construction of non--local conserved
charges, which can be shown to satisfy the commutation relations of 
a quantum 
affine algebra. We will not go into any detail regarding the
construction of these non--local charges. What is important for the
construction of $S$-matrices is the fact that the construction of
these non-local charges gives us the explicit form of the Lorentz
spins and the dependence of the deformation parameter $q$ on the
coupling constant $\beta$. It was found that that the Lorentz spins of
the conserved charges in imaginary ATFTs have the form
\be
s_i = \frac{8\pi}{\beta^2\al_i^2} - 1\;, \hs{1cm} (\forall
i=0,1,\dots,n)\;. \label{atftspins}
\ee
Note in particular that for the case of simply-laced algebras, in
which all simple roots have the same length, the Lorentz spins are all
equal and the $R$-matrix in the spin gradation therefore is equal to
the $R$-matrix in the principal gradation.   
Using (\ref{atftspins}) it is possible to derive the explicit $\theta$
dependence of the spectral parameter $x(\theta)$  in the homogeneous 
gradation $R$-matrix from equation (\ref{xth1}) in which the $n_i$'s
are now the Kac marks of the dual algebra ${\hat g}^{\vee}$. In
\cite{gande95b} we found the result
\be
x(\th) = \exp\(\frac{4\pi h}{\beta^2} - h^{\vee}\)\th =
\exp(h\la\th)\;, \label{xth2} 
\ee
in which we have introduced the coupling constant dependent
function\footnote{This function $\la(\beta)$ is the analogue
     of the $B(\beta)$ used in the $S$-matrices for real coupling
     ATFTs \cite{brade90}. Note, however, that $\la(\beta)$ is
     computed exactly using the quantum affine symmetry algebra,
     whereas $B(\beta)$ was only conjectured and checked to low orders
     in perturbation theory.}   
\be
\la \equiv \frac{4\pi}{\beta^2} - \frac{h^{\vee}}h\;.
\ee 
In the remainder of this paper we always assume that $\beta$ takes
values such that $\la$ is positive.

The other important information which we extract from the paper by
Felder and LeClair is the form of the deformation parameter $q$.
It is possible to deduce that 
\be
q = \exp\left(\frac{4 \pi^2 i}{\beta^2}\right)\;. \label{qbeta}
\ee
In particular we notice that in the case of self--dual algebras $q =
-\exp(i\pi\la)$, from which it emerges that $S$-matrix poles at $x =
q^k$ (for integer $k$) do not depend on the coupling constant $\beta$.
This reflects the fundamental property that the mass ratios of the
solitons in the self--dual cases remain constant under
renormalisation \cite{macka95}.  

{}From the study of the properties of classical soliton solutions of
ATFTs we also know that the topological charges of the solitons lie in
the weight spaces of the fundamental representations of $g$. Together
with the assumption of a quantum affine symmetry this observation
leads us to expect that the quantum affine Toda solitons can be
grouped into mass degenerate multiplets $V_1,V_2,\dots,V_n$
corresponding to the $n$ fundamental representations
of \uq{{\hat g}^{\vee}}, in which $n$ is the rank of $g$. 
Thus a two particle $S$-matrix describing the scattering of affine
Toda solitons should be of the form (\ref{genSM}) in which the
$R$-matrix gradation is determined by (\ref{atftspins}).

%
%
\section{The $a_n^{(1)}$ invariant $S$-matrices}

\thinlines

In this section the general scheme for constructing exact $S$-matrices
{}from trigonometric \mbox{$R$-matrices} will be applied to the $R$-matrices
associated with the fundamental representations of \uq{\ao}. 
The construction of the soliton--soliton $S$-matrices in section 2.1
and 2.2 essentially  reviews Hollowood's construction of the
$S$-matrices for the scattering of $\ao$ affine Toda solitons in
\cite{hollo90,hollo93}, the only new result being an explanation of
the inclusion of the minimal Toda factor. We then study the pole
structure of these $S$-matrices and compute the scattering amplitudes
for the scattering of bound states. 

\subsection{$R$-matrix fusion and crossing properties}\label{ssecanrmf}

We may write the \uq{\ao} invariant $R$--matrices in the following form:
\be
\cR_{a,b}(x) = \sum_{c=0}^{\min(b,n+1-a)}\prod_{i=1}^c\lag
2i+a-b\rag \cP_{\la_{a+c}+\la_{b-c}}\;, \label{an1rm}
\ee
in which we have labelled the projectors $\cP$ by the highest weights
of the associated irreducible modules. $\la_1,\la_2,...,\la_n$ are the
highest weights of the fundamental modules $V_1,V_2,...,V_n$ and we
set $\la_0=\la_{n+1}=0$. We have also used the bracket notation
\be
\lag a \rag \equiv \frac{1-xq^a}{x-q^a}\;.
\ee 
The $R$-matrix (\ref{an1rm}) was given in this form in \cite{hollo94}
and also in \cite{deliu94}. 

For our purpose, the first important feature of these $R$-matrices is
their fusion properties. 
We are always free to rescale $R$-matrices by a scalar factor and we
have chosen our $R$-matrix normalisation such that the projector
$\cP_{\la_a+\la_b}$ always has coefficient $1$. However, this form of
the $R$-matrix is not the one preserved by fusion.
Let us denote the $R$-matrix preserved by fusion by
$\cR^{\prime}_{a,b}(x)$, which means    
\be
\cR'_{a,b}(x) = \prod_{j=1}^a \prod_{k=1}^b
\lb \cR'_{1,1}(xq^{-2-a-b+2j+2k}) \rb_{a+1-j,a+k}\;, \label{rprime}
\ee
in which the equation acts on $V_a\ot V_b \subset V_1^{\ot (a+b)}$
and $[\;]_{j,k}$ indicates that the $R$-matrix is taken to act
on the $j$th and $k$th $V_1$'s. The product is taken in order of
increasing $j$ and $k$. To make this more clear we could also write this
in the form
\bea
\cR^{\prime}_{a,b+c}(x) &=& \left[I_b\ot
\cR^{\prime}_{a,c}(xq^b)\right]  
\left[\cR^{\prime}_{a,b}(xq^{-c})\ot I_c\right]\;, \nn \\ 
\cR^{\prime}_{a+b,c}(x) &=& \left[\cR^{\prime}_{a,c}(xq^b)\ot
I_b\right] \left[I_a\ot\cR^{\prime}_{b,c}(xq^{-a}\right]\;, \nn
\eea
in which the first equation is restricted to $V_a\ot V_{b+c}$ and the
second to $V_{a+b}\ot V_c$. 
Since the totally antisymmetric representation (i.e. $V_{\la_{a+b}}$)
is obtained by fusing totally antisymmetric representations only,
the $R$-matrices $\cR^{\prime}_{a,b}$ must be those in which the
projector $\cP_{\la_{a+b}}$ has prefactor $1$ (or in the case of $a+b
\geq n+1$ those in which $\cP_{\la_{a+b}}$ would have prefactor $1$,
if the tensor product graph was not truncated).  Therefore the
$R$-matrices $\cR^{\prime}_{a,b}$  are related to $\cR_{a,b}$ in the
following way:  
\be
\cR_{a,b}(x) = \prod_{k=1}^b \, \lag a-b+2k\rag \;
\cR'_{a,b}(x)\;. \label{rrconnect}
\ee
Let us define a scalar factor $k_{a,b}(x)$ such that
\[
k_{a,b}(x) \cR_{a,b}(x) \equiv \prod_{j=1}^a \prod_{k=1}^b
\lb \cR_{1,1}(xq^{-2-a-b+2j+2k}) \rb_{a+1-j,a+k}\;. 
\]
Using (\ref{rprime}) and (\ref{rrconnect}) we can compute $k_{a,b}(x)$
explicitly and obtain  
\[
k_{a,b}(x) = (-1)^a \prod_{k=1}^b \lag a+b-2k\rag\;.
\]
Following (\ref{xth2}) and (\ref{qbeta}) we now set
\be
x = e^{2i\pi\mu} \hs{1cm} \mbox{and} \hs{1cm} q = e^{i\pi\la}\;,
\label{xqthbeta} 
\ee
in which $\mu$ and $\la$ are related to the rapidity and coupling
constant\footnote{Note, however, that we follow
\cite{gande96,gande96c} and choose $q$ to differ from (\ref{qbeta}) by
a minus sign. This has no effect on the $S$-matrix pole structure and
agrees with the conventions in \cite{hollo93}.}:
\be
\mu = -i\frac{h\la}{2\pi}\th\;,\hs{1.5cm} \la =
\frac{4\pi}{\beta^2}-1\;. \label{anmula}
\ee
And $h$ denotes the Coxeter number of the underlying affine Lie
algebra, which for the case of $\ao$ is given by
\be 
h=n+1\;.
\ee
In terms of these variables we can write
\be
k_{a,b}(x(\mu)) = (-1)^{a+b}
\prod_{k=1}^b \frac{\sin\(\pi(\mu+\frac{\la}2(a+b-2k))\)} 
{\sin\(\pi(\mu-\frac{\la}2(a+b-2k))\)}\;. \label{an1kab}
\ee
\vs{0.1cm}

\noi In order to proceed also we need the following crossing formula for
the $R$-matrix: 
\be 
\cR_{1,1}(x) = 
\frac{\sin(-\pi\mu)}{\sin(\pi(\mu-\la))}
\cR_{1,1}^{(cross)}(x^{-1}q^{n+1})\;, \label{an1rmcrossing}  
\ee
in which the `crossed' $R$-matrix is given by
\be 
\cR_{1,1}^{(cross)}(x) = (C_n\ot
I_1)\left[\si_{1n}\cR_{n,1}(x)\right]^{t_1}\si_{1n}(I_1\ot
C_1)\;. 
\ee
The formula (\ref{an1rmcrossing}) has been derived by Hollowood in 
\cite{hollo90}.

\subsection{The $S$-matrix scalar factor}\label{ssecansf}

In order to use the above $R$-matrices as scattering matrices we
must multiply them by an overall scalar factor which ensures unitarity
and crossing symmetry. We therefore make the following ansatz:
\be
S_{a,b}(\th) = F_{a,b}(\mu)k_{a,b}(\mu) \tau
\cR_{a,b}(x) \tau^{-1}\;, \label{an1sm}
\ee
in which $\cR_{a,b}(x)$ and $k_{a,b}(\mu)$ are as given above and
$\tau$ denotes the gauge transformation from the homogeneous to the
principal gradation\footnote{Recall that the physically relevant spin
gradation is equal to the principal gradation in the case of
self--dual algebras.}. The overall scalar factor $F_{a,b}(\mu)$
will be constructed in the following. 

We have chosen the definition of $x$ and $q$ in such a
way that the poles in the $S$-matrix, which correspond to the fusion of
two solitons of type $a$ and $b$ into a soliton of type $a+b$, are at
$x=q^{a+b}$. Therefore, $S_{a,b}(\th)$ satisfies the same fusion
formula as $\cR_{a,b}$ and we therefore must have
\be
F_{a,b}(\mu) = \prod_{j=1}^a \prod_{k=1}^b
F_{1,1}(\mu+\frac{\la}2(a+b-2j-2k+2))\;.\label{anFfusion}
\ee
Thus all $F_{a,b}(\mu)$ can be computed from the lowest factor
$F_{1,1}(\mu)$ via this fusion formula, and it is therefore sufficient
to determine the factor $F_{1,1}(\mu)$.  
Putting the ansatz (\ref{an1sm}) (with $a=b=1)$ into equation
(\ref{unitarity}) and using the $R$-matrix normalisation 
(\ref{rmnormalisation}), we obtain
\be
F_{1,1}(\mu)F_{1,1}(-\mu) = 1\;, \label{an1iter1} 
\ee
since $k_{1,1}(\mu) = 1$. 
In order to obtain the condition imposed on $F_{1,1}(\mu)$ by the
requirement of $S$-matrix crossing symmetry, we use the second line of
equation (\ref{crossing}) with $a = b = 1$ and the crossing property
(\ref{an1rmcrossing}) of the $R$-matrix. We then obtain
\be 
F_{1,1}(\mu) = F_{n,1}(-\mu+\frac{n+1}2\la)
k_{n,1}(-\mu+\frac{n+1}2\la)
\frac{\sin(\pi(\mu-\la))}{\sin(-\pi\mu)}\;. 
\ee
Noting that $k_{n,1}(-\mu+\frac{n+1}2\la) =
(-1)^{n+1}\frac{\sin(\pi(\mu-n\la))} 
{\sin(\pi(\mu-\la))}$ and using the
fusion formula for $F_{n,1}(\mu)$ we obtain
\be
F_{1,1}(\mu) (-1)^n \frac{\sin(\pi\mu)}{\sin(\pi(\mu-n\la))} =
\prod_{k=1}^n F_{1,1}(-\mu+k\la)\;. \label{an1iter2}
\ee
Combining (\ref{an1iter1}) and (\ref{an1iter2}) we finally arrive at
\be
\prod_{k=0}^n F_{1,1}(-\mu+k\la) = (-1)^n
\frac{\sin(\pi\mu)}{\sin(\pi(\mu-n\la))}\;. \label{a1iter1}
\ee
One solution to this equation has been found by Hollowood in
\cite{hollo93}\footnote{Note that the equation $(B.1)$ in appendix B
	of \cite{hollo93} is the same as equation
	(\ref{a1iter1}) if $\mu$ is not a negative integer.}. 
There is, however, an infinite number of different solutions. We
are only interested in the solution with a minimum number of poles and
zeros on the physical strip ($0\leq\mu\leq \frac{n+1}2\la$).
For later convenience let us remove the overall minus sign from
the right hand side of (\ref{a1iter1}) and write
\be
\prod_{k=0}^n -F_{1,1}(-\mu+k\la) = 
\frac{\sin(-\pi\mu)}{\sin(\pi(\mu-n\la))}\;. \label{a1iter}
\ee
Hence, we wish to solve an equation of the general form
\be
\prod_{k=0}^n {\cal F}(-\mu+k\la) = {\cal C}(\mu)\;, \label{Fequ}
\ee
for a given function ${\cal C}(\mu)$.
One can immediately write down two (formal) solutions to this equation 
\bea
{\cal F}_1(\mu) &=& \prod_{j=1}^{\infty} \frac{{\cal C}(-\mu+jh\la-\la)} 
{{\cal C}(-\mu+jh\la)}\;, \label{F1equ} \\
{\cal F}_2(\mu) &=& \prod_{j=1}^{\infty} \frac{{\cal C}(-\mu-(j-1)h\la)}  
{{\cal C}(-\mu-(j-1)h\la-\la)}\;. \label{F2equ}
\eea
It is easy to see that these functions solve
equation (\ref{Fequ}) if $h=n+1$, which is the Coxeter number of $\ao$.
Of course these solutions only make
sense if we can prove the convergence of the infinite products. 

If we simply solve equation (\ref{Fequ}) with ${\cal C}(\mu)$
equal to the right hand side of (\ref{a1iter}) we will find that
solutions ${\cal F}_1$ and ${\cal F}_2$ both have an infinite number
of poles in the physical strip. We can, however, find a scalar factor
with only a finite number of poles, if we combine the two solutions in a
certain way. In order to do this 
let us write the right hand side of equation (\ref{a1iter}) in
terms of Gamma functions
\be
\frac{\sin(-\pi\mu)}{\sin(\pi(\mu-n\la))} = 
\frac{\Ga(\mu-n\la) \Ga(1-\mu+n\la)} {\Ga(-\mu) \Ga(1+\mu)}\;.
\ee
We notice that we would at most obtain a finite number of zeros and poles
on the physical strip if in the infinite product only terms of the
form $\Ga(\pm \mu + jh\la \pm \dots)$ appeared and no terms 
with a minus sign in front of $jh\la$ appeared. (Recall that $\la > 0$.)
It is then simple to construct a solution with this property, if
we define  
\be
{\cal C}_1(\mu) \equiv \frac{\Ga(\mu-n\la)}{\Ga(1+\mu)} \hs{0.5cm}
\mbox{and} \hs{0.5cm} 
{\cal C}_2(\mu) \equiv \frac{\Ga(1-\mu+n\la)}{\Ga(-\mu)} \;.
\ee
Now we use (\ref{F1equ}) for ${\cal C}_1$ and (\ref{F2equ}) for ${\cal
C}_2$ and we obtain the desired solution to equation (\ref{a1iter}) as
the product of these two solutions:
\bea
F^{(A)}(\mu) &=& \prod_{j=1}^{\infty} \frac{\Ga(\mu+jh\la-n\la)
\Ga(\mu+jh\la-\la+1)} {\Ga(-\mu+jh\la-n\la)\Ga(-\mu+jh\la-\la+1)} \nn
\\ && \hs{9pt} \times \frac{\Ga(-\mu+jh\la-n\la-\la+1)
\Ga(-\mu+jh\la)} 
{\Ga(\mu+jh\la-n\la-\la+1) \Ga(\mu+jh\la)}\;. \label{Fminus}
\eea
This is the solution Hollowood found in
\cite{hollo93}\footnote{The different prefactor in front of the
infinite product is due to the different $R$-matrix normalisation.}. 
However, with our method we are able additionally to obtain
a very similar but different solution by simply rewriting the
right hand side of (\ref{a1iter}) in the following trivial way
\be
\frac{\sin(-\pi\mu)} {\sin(\pi(\mu-n\la))} =
\frac{\sin(\pi\mu)} {\sin(\pi(-\mu+n\la))} =
\frac{\Ga(-\mu+n\la) \Ga(1+\mu-n\la)} {\Ga(\mu)
\Ga(1-\mu)}\;. 
\ee
Using the same procedure as above (now with  ${\cal C}_1 =
\frac{\Ga(1+\mu-n\la)}{\Ga(\mu)}$ and  ${\cal C}_2 =
\frac{\Ga(-\mu+n\la)}{\Ga(-\mu)}\;$)
we obtain another solution to
(\ref{a1iter}) in the form
\bea
F^{(B)}(\mu) &=& - \prod_{j=1}^{\infty}
\frac{\Ga(\mu+jh\la-n\la+1) 
\Ga(\mu+jh\la-\la)} {\Ga(-\mu+jh\la-n\la+1)\Ga(-\mu+jh\la-\la)} \nn
\\ && \hs{20pt} \times \frac{\Ga(-\mu+jh\la-n\la-\la+1) \Ga(-\mu+jh\la)}
{\Ga(\mu+jh\la-n\la-\la+1) \Ga(\mu+jh\la)}\;.  \label{an1scalar}
\eea
Thus we have obtained two slightly different solutions. In
\cite{hollo93} Hollowood found that the factor $F^{(A)}$ does not
quite give the correct pole structure. It was thus necessary to
include an additional CDD--factor in the $S$-matrix conjecture. This
CDD-factor was equal to the minimal $a_n$ Toda $S$-matrix. For $a=b=1$
this minimal Toda $S$-matrix is simply 
\be
S_{1,1}^{(min)}(\mu) = \frac{\sin(\frac{\pi}{h\la}(\mu+\la))}
{\sin(\frac{\pi}{h\la}(\mu-\la))} \;.
\ee
Using the expansion of the sine function in terms of an infinite
product we can rewrite this in the following form: 
\[
S_{1,1}^{(min)}(\mu) = \frac{\mu+\la}{\mu-\la} \prod_{j=1}^{\infty}
\frac{(jh\la)^2 - (\mu+\la)^2} {(jh\la)^2 - (\mu-\la)^2} = 
-\prod_{j=1}^{\infty} \frac{(\mu+jh\la-n\la)(-\mu+jh\la-\la)}
{(\mu+jh\la-\la)(-\mu+jh\la-n\la)}\;.  
\]
Using the fundamental property of the Gamma function we can then see that 
\be
F^{(B)}(\mu) =  S_{1,1}^{(min)}(\mu) F^{(A)}(\mu)\;.
\ee
If we now carefully examine the poles and zeros of $F^{(A)}(\mu)$
and $F^{(B)}(\mu)$ on the physical strip ($0\leq \mu\leq
\frac{n+1}2 \la$), we find that both functions displays simple
poles at $\mu = m$ (for all $m = 1,2,\dots\leq \frac{n+1}2 \la$) and
simple zeros at $\mu = \la + m$ (for all $m = 1,2,\dots\leq
\frac{n-1}2 \la$). $F^{(A)}$, however, additionally 
displays a zero at $\mu = \la$. Thus the 
scalar function originally found in \cite{hollo93} is not minimal and 
the additional pole in the included CDD--factor only serves to cancel
this additional zero.
This derivation explains the inclusion of the minimal Toda $S$-matrix
in \cite{hollo93}. There are in fact two (almost) minimal solutions to
equation (\ref{a1iter}) and they are distinguished by exactly the
minimal $a_n$ Toda $S$-matrix. 

Therefore we choose the solution $F^{(B)}(\mu)$ as our scalar
factor for $S_{1,1}(\th)$. We define $F_{1,1}(\mu) \equiv
F^{(B)}(\mu)$ and all scalar factors for the higher $S$-matrices
can then be determined by the fusion formula (\ref{anFfusion}). After a
somewhat lengthy but straightforward calculation, we find the
following expression for the general scalar factor $F_{a,b}(\mu)$:
\bea
F_{a,b}(\mu) &=& (-1)^{a+b}\; \prod_{k=1}^{b-1}\;
\mbox{\large $\frac{\sin\(\pi\(\mu+\frac{\la}2(-a+b-2k)\)\)}
{\sin\(\pi\(\mu+\frac{\la}2(a+b-2k)\)\)}\;  
\frac{\sin\(\frac{\pi}{h\la}\(\mu+\frac{\la}2(a+b-2k)\)\)}
{\sin\(\frac{\pi}{h\la}\(\mu+\frac{\la}2(-a+b-2k)\)\)}$} \nn \\ 
&& \hs{34pt} \times \prod_{j=1}^{\infty}
\mbox{\Large $\frac{\Ga\(\mu+jh\la-\frac{\la}2(2n+2-a-b)+1\)
\Ga\(\mu+jh\la-\frac{\la}2(a+b)\)}
{\Ga\(-\mu+jh\la-\frac{\la}2(2n+2-a-b)+1\) 
\Ga\(-\mu+jh\la-\frac{\la}2(a+b)\)}$} \nn \\
&& \hs{54pt} \times \mbox{\Large
$\frac{\Ga\(-\mu+jh\la-\frac{\la}2(2n+2+a-b)+1\)
\Ga\(-\mu+jh\la-\frac{\la}2(-a+b)\)}
{\Ga\(\mu+jh\la-\frac{\la}2(2n+2+a-b)+1\)
\Ga\(\mu+jh\la-\frac{\la}2(-a+b)\)}$}\;. \nn \\ \label{anFab}
\eea

As mentioned earlier, it is of course important to show that these
infinite products converge at least within the physical strip. This
can be done be rewriting the infinite products of Gamma functions in
terms of integrals over hyperbolic functions. This has been shown in
detail in \cite{gande96c} and for the case of the $\ao$ scalar factor
it was found that
\be 
F_{1,1}(\mu) = \exp \bl \int_0^{\infty} \frac{dt}t 2\sinh(\mu t)
{\cal I}(t) \br \;, \label{integral}
\ee
in which
\be
{\cal I}(t) \equiv
\frac{\sinh(\frac{\la}{2} t) \sinh((\frac n2 \la - \frac12)t)}
{\sinh(\frac{t}2) \sinh((\frac{n+1}2\la t)} \;. 
\ee
A similar form of the $S$-matrix scalar factors for the cases of 
simply laced algebras has recently been constructed in a very
interesting paper \cite{johns96}, which uses a so-called regularised
quantum dilogarithm. This form has 
the advantage that the $S$-matrices can be compared easily with the
time delays in the 
classical scattering of affine Toda solitons. Although these kinds of
integral representations provide a more compact way of writing the
overall scalar factors, for our purposes it is advantageous 
to use infinite products of Gamma functions, since in this case
the study of the pole structure is greatly simplified. 

The $S$-matrix (\ref{an1sm}) with the scalar factor (\ref{anFab}) is
exactly that suggested in \cite{hollo93}. If we take the
underlying algebra to be $a_1^{(1)}$ then this $S$-matrix is the well
known $S$-matrix for the Sine-Gordon solitons \cite{zamol79}. In
\cite{gande95} this \mbox{$S$-matrix} was used to construct the
$S$-matrices for the scattering of bound states in $a_2^{(1)}$
ATFT. In following section we extend this to the general case of
$\ao$. 

\subsection{The bound states} \label{secanbs}

The $S$-matrices (\ref{an1sm}) describe the quantum scattering of a
set of quantum states (which will be referred to as solitons)
grouped into mass degenerate multiplets $V_a$ (for
$a=1,2,...,n$) corresponding to the $n$ fundamental representations of
\uq{\ao}. Let us denote the solitons in a multiplet $V_a$ by $A^{(a)}$. 
Although in the following we will always speak of a `soliton' 
$A^{(a)}$ we should point out that $A^{(a)}$ actually
denotes an entire multiplet $V_a$ of solitons transforming under the
$a$th fundamental representation. The single solitons should therefore 
carry an additional multiplet label, i.e.\ $A^{(a)}_{(j)}$ in which 
$j=1,2,...,\mbox{dim}V_a$. However, in order to construct bound state
$S$-matrices we will not need to distinguish different solitons of the
same multiplet and we will therefore supress the multiplet label in
the following. 

The quantum masses of these solitons are determined by the pole
structure of the $S$-matrices. 
We find that the fusion of two solitons $A^{(a)}$ and $A^{(b)}$ into a
soliton $A^{(a+b)}$ corresponds to the simple pole in the $S$-matrix
at $\mu = \frac{a+b}2 \la$, which is at  $x=q^{a+b}$. 
According to the mass formula (\ref{mass}) the quantum masses of
the elementary solitons must have the form 
\be
M_a = 2mC \sin\(\frac{\pi a}{n+1}\)\;, \hs{1cm}
(a=1,2,\dots,n)\;,\label{ansolmass}  
\ee
with an unspecified overall factor $C$. As we will discuss at the end
of this chapter, these masses are proportional to the masses of the
classical solitons in $\ao$ ATFTs.
We can also see that the solitons in the multiplets $V_a$ and
$V_{n+1-a}$ have the same mass. They are charge conjugate to each
other and transform into each other under time reversal. We therefore
introduce the following notation for a `charge conjugate' soliton: 
\[
\ol A^{(a)}(\th) \equiv A^{(n+1-a)}(\th)\;.
\]  

\subsubsection{Breather bound states}

It has already been shown that in Sine--Gordon theory as well as in
the $\ato$ ATFT two solitons of conjugate type can fuse into
so--called breather bound states. These are bound states with zero
topological charge, which means that they transform under the singlet
representation.  
The poles in the \mbox{$S$-matrix} at which two solitons fuse
into a breather bound state are distinguished by the fact that the
$S$-matrix projects onto the module of the singlet representation at
these poles. Thus, if we denote possible breather poles by $\th_p$,
then we must have  
\be
S_{a,b}(\th_p) \sim \cP_0\;. 
\ee
Now let us look at possible scalar bound states in the $S$-matrices
(\ref{an1sm}).  We first notice that this last equation can only be
true if $a = \ol b$ (in which $\ol b \equiv n+1-b$) since
only in this case does the projector $\cP_0$ appear in the spectral
decomposition of the $\ao$ \mbox{$R$-matrix} (\ref{an1rm}). Therefore,
breathers must be bound states of two solitons of conjugate
type, say $b$ and $\ol b$. From the spectral
decomposition of the corresponding $R$-matrix we can see that possible
breather poles must be contained in the factor
$\lag n+1\rag$, since this is the only factor which appears in front
of the projector $\cP_0$ and in front of no other projector.  
$\lag n+1\rag$ becomes singular at $x=q^{n+1}$ and therefore
\[
\cR_{\ol b,b}(q^{n+1}) \sim \cP_0 \;.
\]
{}From the definitions (\ref{anmula}) we see that $x=q^{n+1}$
corresponds to 
\be 
\mu = \frac{n+1}2\la + m\;, \hs{1cm} (\forall m \in Z)\;,
\label{anbrpol1} 
\ee
of which only the poles with $m = 0,-1,-2,\dots \geq -\frac{n+1}2 \la$
lie in the physical strip.
Of course we also have to examine possible pole--zero cancellations
{}from the overall scalar factor $F_{\ol b,b}(\mu)k_{\ol
b,b}(\mu)$. Using (\ref{an1kab}) and (\ref{anFab}) we find
that no physical strip poles of the form (\ref{anbrpol1}) appear in
this factor, but we do find a simple zero at $\mu=\frac{n+1}2\la$,
which cancels the corresponding pole for $m=0$ in (\ref{anbrpol1}). We
are therefore left with the following simple poles in the $S$-matrix: 
\[ 
\th_p = i\pi\(1-\frac{2p}{(n+1)\la}\)\;,
\] 
or in terms of $\mu$
\be
\mu_p \equiv \mu(\th_p) = \frac{n+1}2\la - p\;, 
\ee
in which $p = 1,2,\dots \leq \frac{n+1}2\la$. We conjecture that
these poles correspond to the fusion of two solitons into bound
states as depicted in {\em figure 1}. 
These bound states transform under the singlet representation
and we will call them breathers. Let us denote the breather bound
states of two solitons of type $b$ and $\ol b$ by the symbols
$B_p^{(b)}(\th)$. We can also define the notion of a `conjugate
breather', in which the order of the incoming solitons of 
type $b$ and $\ol b$ is reversed, and we introduce the notation
$\ol B_p^{(b)}(\th) \equiv B^{(n+1-b)}_p(\th)$.
The breathers carry an additional quantum number $p$, the so--called
excitation number, which takes integer values
$p=1,2,...\leq\frac{n+1}2 \la$. Thus the number of breather states in
the spectrum decreases with increasing coupling constant $\beta$ and
for $\la < \frac{2}{n+1}$ (i.e.\ $\beta^2 > 4\pi\frac{n+1}{n+3}\;$) 
all breather states disappear from the spectrum.

\vspace{0.5cm}
\begin{center}
\begin{picture}(120,120)(-10,-10)
%
%
\put(0,0){\line(1,1){40}}
\put(40,40){\line(1,-1){40}}
\put(40,40){\line(0,1){64}}
\put(40,40){\circle{16}}
\put(40,17){\vector(0,1){14}}
\put(22,58){\vector(1,-1){10}}
\put(58,58){\vector(-1,-1){10}}
\put(24,7){\shs{\scs{$\frac{n+1}2 \la-p$}}}
\put(-4,65){\shs{\scs{$\frac p2 +\frac{n+1}4\la$}}}
\put(50,65){\shs{\scs{$\frac p2+\frac{n+1}4\la$}}}
\put(-5,-10){\shs{\fns{$\ol A^{(b)}$}}}
\put(76,-10){\shs{\fns{$A^{(b)}$}}}
\put(35,108){\shs{\fns{$B^{(b)}_p$}}}

\put(-15,-34){\shs{\em Figure 1: Breather fusion}}

\end{picture}
\end{center}
\vs{1.5cm}

\noi From the mass relation (\ref{mass}) we
can determine the quantum masses of the breathers
\be 
m_{B_p^{(b)}} = 2M_b\sin\(\frac{\pi p}{(n+1)\la}\)\;, \hs{1.5cm}
(b=1,2\dots,n)\;, 
\ee     
in which $M_b$ is the mass of the $b$th elementary soliton given by
(\ref{ansolmass}).  

\subsubsection{Excited solitons}\label{ssecanes}

By analogy with the study in \cite{gande95} we expect the theory to
contain some sort of excited solitons which are soliton bound states
with non--zero topological charge.
{}From the classical examination of the $\ao$ bound states in
\cite{harde95}, we expect bound states of two solitons of the
same species to exist. Let us therefore look at the $R$-matrix
associated to the tensor product $V_a \ot V_a$ for some $a\in
\{1,2,\dots,n\}$. From (\ref{an1rm}) we get 
\[
\cR_{a,a}(x) = \sum_{c=0}^{\min(a,n+1-a)} \prod_{i=1}^c \lag 
2i\rag \cP_{\la_{a+c}+\la{a-c}} 
= \cP_{2\la_a} +\dots + \prod_{i=1}^{\min(a,n+1-a)} \lag 2i
\rag \cP_{\la_{\min(2a,2a-n-1)}}\;.
\]
We expect the excited solitons to transform under fundamental
representations and we therefore have to consider the poles in the
factor $\lag 2\min(a,n+1-a)\rag$ at which the \mbox{$R$-matrix}
projects onto the fundamental module $V_{\min(2a,2a-n-1)}$. 
After a careful study of the corresponding poles and zeros in the
overall scalar factor $F_{a,a}(\mu)k_{a,a}(\mu)$ we find all poles
at which the soliton $S$-matrix projects onto $V_{\min(2a,2a-n-1)}$. 
We have to distinguish the following three cases:

\noi{\bf i) $a < \frac{n+1}2$} \vs{3pt}\\
The excited soliton poles are contained in the factor $\lag
2a\rag$ which has poles at $x = q^{2a}$, and the overall scalar factor
$F_{a,a}(\mu)k_{a,a}(\mu)$ displays simple zeros on the
physical strip at $\mu=a\la+m$ for $m=1,2,\dots$. Therefore, we end up
with the following simple poles on the physical strip:
\be 
\mu = a\la-p\;, \hs{1.5cm} (\mbox{for } p = 0,1,2,\dots \leq a\la)\;,
\label{anexsolpole1} 
\ee  
which correspond to a fusion process $A^{(a)}+A^{(a)} \to
A^{(2a)}_p$. 
\vs{3pt}

\noi{\bf ii) $a > \frac{n+1}2$} \vs{3pt} \\
Here the excited soliton poles are contained in $\lag
2(n+1-a)\rag$ which displays simple poles at $\mu = (n+1-a)\la +m$ for
integer $m$. However we find corresponding zeros in the scalar factor
for all positive $m$ and we thus end up with the following simple poles
on the physical strip:
\be 
\mu = (n+1-a)\la-p\;, \hs{1.5cm} (\mbox{for } p = 0,1,2,\dots \leq
(n+1-a)\la)\;, 
\ee  
which correspond to the fusion process $A^{(a)}+A^{(a)} \to
A^{(2a-n-1)}_p$.
\vs{3pt}

\noi{\bf iii) $a = \frac{n+1}2$} \vs{3pt} \\
In this case we have $\cR_{a,a}(\th) = \cR_{a,\ol a}(\th)$ and the
excited solitons are nothing other than the breathers $B^{(a)}_p$ which
were studied above.
\vs{3pt}

Thus we have found bound states of two elementary solitons of the same
species which transform under the fundamental representations
$\pi_{2a\,\mbox{\scs{mod}}h}$. We have denoted these excited solitons
by $A^{(2a\,\mbox{\scs{mod}}h)}_p$, in which $p$ is the excitation
number taking values $p = 1,2,\dots,\leq \min(a,n+1-a)\la$. Thus, as
in the case of the breather bound states, the number of excited
solitons in the spectrum of the theory is restricted by the coupling
constant. The quantum masses of these excited solitons are
\bea
\hs{-30pt} m_{A^{(2a)}_p} = 2M_a \cos\(\frac{\pi}{n+1}(a-\frac{p}{\la})\)\;,
\hs{0.4cm}  &&(\mbox{for}\hs{9pt} a = 1,2,\dots <\frac{n+1}2 \nn \\
&& \mbox{and}\hs{9pt} p = 0,1,2,\dots\leq a\la)\;, \nn
\eea
\bea
m_{A^{(2a-n-1)}_p} = 2M_a \cos\(\frac{\pi}{n+1}(a+\frac{p}{\la})\)\;,
\hs{0.4cm}  &&(\mbox{for}\hs{9pt} \frac{n+1}2 < a < n+1 \nn \\
&& \mbox{and}\hs{9pt}  p = 0,1,2,\dots\leq (n+1-a)\la)\;. \nn
\eea    
We find that the excited solitons with the lowest mass, i.e.\ $A^{(2a\,
\mbox{\scs{mod}}h)}_0$, are indeed just the elementary solitons $A^{(2a\,
\mbox{\scs{mod}}h)}$ themselves. 

Before we discuss any of the other simple and multiple poles in the
soliton $S$-matrices we will first construct the scattering amplitudes
for the scattering of breathers and excited solitons in the following
subsection.

\subsection{The bound state $S$-matrices}\label{ssecanbsm}

Now we are able to compute the $S$-matrices for the scattering of 
breather bound states by using the bootstrap equations. 
We start with the $S$-matrix  for the scattering of a breather of type
$b$ with an elementary soliton of type $a$ and we therefore have to use
the following bootstrap equation: 
\be
S_{A^{(a)}B_p^{(b)}}(\th)\(I_a\ot P_0\) = 
P_0\ot I_a \;.\; I_b \ot S_{a,\ol b}(\th+\frac12 \th_p)
\;.\; S_{a,b}(\th-\frac12 \th_p)\ot I_{\ol b}\;, \label{anbrbtrapequ}
\ee
in which $P_0$ denotes the projector from $V_b\ot V_{\ol b}$ onto the
singlet space, such that both sides of equation (\ref{anbrbtrapequ})
map $V_a \ot V_{b}\ot V_{\ol b}\;$ into $V_a\;$. This bootstrap equation is
illustrated in {\em figure 2}, in which as usually time is meant to
run upwards. 

%
%
%
\begin{center}
\begin{picture}(240,130)
\put(10,10){\line(1,2){50}}
\put(20,110){\line(1,-2){30}}
\put(50,10){\line(0,1){40}}
\put(90,10){\line(-1,1){40}}
\put(110,60){\line(1,0){10}}
\put(110,64){\line(1,0){10}}
\put(145,10){\line(1,2){50}}
\put(150,110){\line(1,-2){15}}
\put(165,10){\line(0,1){70}}
\put(235,10){\line(-1,1){70}}
\put(0,0){\shs{\fns{$A^{(a)}$}}}
\put(45,0){\shs{\fns{$A^{(b)}$}}}
\put(85,0){\shs{\fns{$A^{(\ol b)}$}}}
\put(15,114){\shs{\fns{$B^{(b)}_p$}}}
\put(55,114){\shs{\fns{$A^{(a)}$}}}
\put(135,0){\shs{\fns{$A^{(a)}$}}}
\put(160,0){\shs{\fns{$A^{(b)}$}}}
\put(230,0){\shs{\fns{$A^{(\ol b)}$}}}
\put(145,114){\shs{\fns{$B^{(b)}_p$}}}
\put(190,114){\shs{\fns{$A^{(a)}$}}}

\put(30,-34){\shs{\em Figure 2: Bootstrap for $S_{A^{(a)}B^{(b)}_p}$}}

\end{picture}
\end{center}
\vs{1.3cm}

\noi Using the crossing symmetry of the $S$-matrix we can write
\bea
S_{A^{(a)}B_p^{(b)}}(\th)\(I_a\ot P_0\) &=& 
P_0\ot I_a \;.\; I_{b} \ot S^{cross}_{a, \ol b}(i\pi-(\th+\frac12
\th_p)) \;.\; S_{a,b}(\th-\frac12 \th_p)\ot I_{\ol b}  \nn \\
&=& {\cal F}(\mu)\lb \cP_0\ot I_a \;.\; I_b \ot
\cR^{cross}_{a,\ol b}(x^{-1}q^{(n+1)/2}) \;.\; \cR_{a,b}(xq^{(n+1)/2})
\ot I_{\ol b} \rb \nn \\ 
&=& {\cal F}(\mu)\(I_a \ot P_0\)\;. \label{a1SAB}
\eea
The last step in this calculation is non-trivial and uses the
$R$-matrix unitarity (a proof of this can be found in the appendix of
\cite{gande96c}).
Thus the $S$-matrix $S_{A^{(a)}B_p^{(b)}}(\th)$ is simply a
scalar factor equal to ${\cal F}(\mu)$ which is given by  
\be
{\cal F}(\mu) = F_{b,a}(\mu(i\pi-\th-\frac12
\th_p))F_{a,b}(\mu(\th-\frac12 \th_p)) k_{b,a}(\mu(i\pi-\th-\frac12 
\th_p))k_{a,b}(\mu(\th-\frac12 \th_p))\;. 
\ee
We can compute this explicitly and we find a surprisingly simple
expression for $S_{A^{(a)}B_p^{(b)}}(\th)$ in terms of a finite product
of sine functions: 
\bea
S_{A^{(a)}B_p^{(b)}}(\th) &=& \hs{11pt} \prod_{l=1}^{p-1} \frac{\lb
\frac{\la}2(a+b-\frac{n+1}2)-\frac p2+l\rb \lb
-\frac{\la}2(a+b+\frac{n+1}2)-\frac p2+l\rb} {\lb
\frac{\la}2(a-b-\frac{n+1}2)-\frac p2+l\rb 
\lb -\frac{\la}2(a-b+\frac{n+1}2)-\frac p2+l\rb} \nn \\
&& \times \prod_{k=1}^b \frac{\lb
\frac{\la}2(a+b-\frac{n+1}2-2k+2)+\frac p2 \rb 
\lb \frac{\la}2(b-a-\frac{n+1}2-2k)-\frac p2\rb} 
{\lb \frac{\la}2(b-a-\frac{n+1}2-2k+2)+\frac p2\rb 
\lb \frac{\la}2(a+b-\frac{n+1}2-2k)-\frac p2\rb}\;,\nn \\
\label{ansmAB} 
\eea
in which we have introduced the following bracket notation: 
\be
\lb y \rb \equiv \sin(\frac{\pi}{h\la}(\mu+y))\;, \label{bracket1}
\ee 
and for later use we also introduce 
\be
\bl y \br \equiv \frac{\lb y\rb}{\lb -y\rb}\;. \label{bracket2}
\ee
It can be easily checked that these $S$-matrix elements are themselves
crossing symmetric, i.e. 
$S_{A^{(a)}B_p^{(b)}}(i\pi-\th) = S_{A^{(a)}B_p^{(b)}}(\th)$ and
therefore satisfy the following required symmetry conditions:
\[
S_{A^{(a)}B_p^{(b)}}(\th) = S_{\ol A^{(a)}\ol B_p^{(b)}}(\th) =
S_{\ol B_p^{(b)}A^{(a)}}(\th) = S_{B_p^{(b)}\ol A^{(a)}}(\th)\;, 
\]
and since $S_{\ol A^{(a)}B_p^{(b)}} = S_{A^{(h-a)}B_p^{(b)}}$, we also
have 
\[ 
S_{B_p^{(b)}A^{(a)}}(\th) = S_{\ol B_p^{(b)}\ol A^{(a)}}(\th) \nn = 
S_{A^{(a)}\ol B_p^{(b)}}(\th) = S_{\ol A^{(a)}B_p^{(b)}}(\th)\;.
\] 
Using these identities we can apply the bootstrap method again in order
to obtain the breather--breather $S$-matrices. 
If we replace the soliton $A^{(a)}$ in {\em figure 2}
with a breather $B^{(a)}_r$, 
we obtain the following bootstrap equation:
\bea
S_{B^{(a)}_rB_p^{(b)}}(\th) &=& S_{A^{(a)}B_p^{(b)}}(\th+\frac12
\th_r) S_{\ol A^{(a)}B_p^{(b)}}(\th-\frac12 \th_r)  \nn \\
&=& \hs{11pt} \prod_{l=1}^{p-1} \bl \frac{\la}2(b-a)
+\frac{r-p}2+l \br \bl \frac{\la}2(a-b) +\frac{r-p}2 +l\br \nn \\
&& \hs{17pt} \times \bl\frac{\la}2(a+b) -\frac{p+r}2+l\br \bl
-\frac{\la}2(a+b)-\frac{p+r}2 +l\br \nn \\ 
&& \times  \prod_{k=1}^b  \bl \frac{\la}2(b-a-2k+2)
+\frac{p+r}2\br \bl \frac{\la}2(a+b-2k) +\frac{r-p}2\br \nn \\
&& \hs{17pt} \times \bl\frac{\la}2(a+b-2k+2) +\frac{p-r}2\br \bl
\frac{\la}2(b-a-2k)-\frac{p+r}2 \br\;. 
\label{anbrbrsm}
\eea
Finally, we also compute the $S$-matrix for the
scattering of an excited soliton with a breather:
\bea
S_{A^{(2a)}_rB_p^{(b)}}(\th) &=& \hs{2pt} \prod_{l=1}^{p-1} \frac{\lb
\frac{\la}2(2a+b-\frac{n+1}2)-\frac{p+r}2+l\rb 
\lb -\frac{\la}2(2a+b+\frac{n+1}2)+\frac{r-p}2+l\rb} 
{\lb \frac{\la}2(2a-b-\frac{n+1}2)-\frac{p+r}2+l\rb 
\lb -\frac{\la}2(2a-b+\frac{n+1}2)+\frac{r-p}2+l\rb} \nn \\
&& \hs{10pt} \times  \frac{\lb
-\frac{\la}2(b+\frac{n+1}2)-\frac{p+r}2+l\rb 
\lb \frac{\la}2(b-\frac{n+1}2)+\frac{r-p}2+l\rb} 
{\lb \frac{\la}2(b-\frac{n+1}2)-\frac{p+r}2+l\rb 
\lb -\frac{\la}2(b+\frac{n+1}2)+\frac{r-p}2+l\rb} \nn \\
&& \hs{-9pt} \times \prod_{k=1}^b \frac{\lb
\frac{\la}2(2a+b-\frac{n+1}2-2k+2)+\frac{p-r}2 \rb 
\lb \frac{\la}2(b-2a-\frac{n+1}2-2k)+\frac{r-p}2\rb} 
{\lb \frac{\la}2(2a+b-\frac{n+1}2-2k)-\frac{p+r}2\rb 
\lb \frac{\la}2(b-2a-\frac{n+1}2-2k+2)+\frac{p+r}2\rb} 
\nn \\ 
&& \hs{10pt} \times \frac{\lb
\frac{\la}2(b-\frac{n+1}2-2k)-\frac{p+r}2 \rb 
\lb \frac{\la}2(b-\frac{n+1}2-2k+2)+\frac{p+r}2\rb} 
{\lb \frac{\la}2(b-\frac{n+1}2-2k+2)+\frac{p-r}2\rb 
\lb \frac{\la}2(b-\frac{n+1}2-2k)+\frac{r-p}2\rb}\;.
\eea
\vs{0.1cm}

The same procedure could in principle be applied again in order to
construct the \mbox{$S$-matrices} for the scattering of two excited
solitons. This $S$-matrix, however, would not be just a scalar
function but, like $S_{a,b}(\th)$, an intertwiner on the tensor
product of the two corresponding modules. This construction
remains beyond the scope of this paper.

\subsection{The spectrum}\label{ssecansp}

We conjecture that the entire spectrum of states of a quantum field
theory, the on-shell information of which is provided by the above
$S$-matrices, consists of fundamental solitons and two kinds of bound
states, the breathers and the excited solitons as described above. 
This is obviously a bold assertion given the fact that the
$S$-matrices contain a vast number of so far unexplained
poles. However, it was already discovered in previous work that a very
large number of poles in trigonometric $S$-matrices can be explained
by higher order diagrams, many of which involve a generalised
Coleman-Thun mechanism (see \cite{colem78,corri93,gande95} for
details).  

Let us, as an example, consider the pole at which the $S$-matrix projects
onto the module of the fundamental representation $\pi_{\la_{a+b}}$.
For the sake of simplicity let us assume that $a\geq b$ and $b\leq
n+1-a$. The $R$-matrix in this case is then given by
\be
\cR_{a,b}(x) = \prod_{c=0}^b\, \prod_{i=1}^c \lag 2i+a-b\rag
\cP_{\la_{a+c}+\la_{b-c}}\;. 
\ee
{}From this formula we can see that at the poles in the factor $\lag
a+b\rag$ the $R$-matrix projects onto $V_{a+b}$. The factor $\lag
a+b\rag$ has simple poles at $x=q^{a+b}$ which correspond to 
$\mu = \frac{\la}2 (a+b) + m$ (for $m\in {Z}$). 
If we carefully study the corresponding zeros and poles in
$F_{a,b}(\mu)k_{a,b}(\mu)$ we find that $S_{a,b}(\mu)$ displays simple
poles on the physical strip at  
\be
\mu = \frac{\la}2 (a+b) - p \;, \hs{1.5cm} \mbox{for } p = 0,1,\dots
\leq \frac{\la}2 (a+b)\;. \label{anCT1pole}
\ee
As mentioned earlier, the pole with $p=0$ among these corresponds to
the fusion process $A^{(a)} + A^{(b)} \to A^{(a+b)}$. The poles with
$p>0$ do not correspond to a fusion process into a bound state
but they correspond to the crossed box process depicted in {\em figure
3}.      

%
%
%
\begin{center}
\begin{picture}(140,160)
\put(0,0){\line(1,2){20}}
\put(120,0){\line(-1,2){20}}
\put(20,40){\line(0,1){60}}
\put(20,40){\line(4,3){80}}
\put(20,100){\line(4,-3){80}}
\put(100,40){\line(0,1){60}}
\put(20,100){\line(-1,2){20}}
\put(100,100){\line(1,2){20}}
\put(60,70){\circle*{8}}
\put(-8,-10){\shs{\fns{$A^{(a)}$}}}
\put(115,-10){\shs{\fns{$A^{(b)}$}}}
\put(115,142){\shs{\fns{$A^{(a)}$}}}
\put(-5,142){\shs{\fns{$A^{(b)}$}}}
\put(4,65){\shs{\tiny{$A^{(b)}$}}}
\put(102,65){\shs{\tiny{$A^{(b)}$}}}
\put(35,46){\shs{\tiny{$A^{(a-b)}$}}}
\put(67,46){\shs{\tiny{$B^{(b)}_p$}}}
\put(36,89){\shs{\tiny{$B^{(b)}_p$}}}
\put(64,89){\shs{\tiny{$A^{(a-b)}$}}}
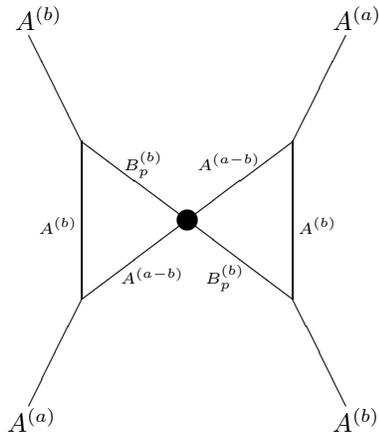
\put(-65,-34){\shs{\em Figure 3: A generalised Coleman 
Thun process in $\ao$}} 
\end{picture}
\end{center}
\vs{1cm}

\noi If the incoming solitons in this diagram have a
rapidity difference equal to the poles (\ref{anCT1pole}) then
the scattering process in the center of the diagram occurs at exactly
$\mu = \frac{n+1}4 \la +\frac a2 \la +\frac p2$. From formula
(\ref{ansmAB}) we can see that the $S$-matrix element
$S_{A^{(a-b)}B^{(b)}_p}$ has a simple zero at this value of rapidity
difference. This zero reduces the expected double pole to the single
poles (\ref{anCT1pole}).  Thus we have shown that the poles
(\ref{anCT1pole}) do not correspond to new bound states, but can be
explained by a generalised Coleman-Thun mechanism.
We also note that the process in {\em figure 3} can only 
exist for the case of $a \neq b$, because of the occurrence of the
soliton $A^{(a-b)}$. This is consistent with the fact that only in the
case $a=b$ do the poles (\ref{anCT1pole}) correspond to the fusion
into an excited soliton.

We have examined a large number of simple and multiple poles and found
that all of these poles can be explained in a similar fashion in terms of
the conjectured spectrum of solitons, breathers and excited
solitons\footnote{For a detailed discussion of a large number of poles
in the case of the $a_2^{(1)}$ theory see \cite{gande95}.}. We
believe that this will prove true for all poles in the soliton and
bound state $S$-matrices and that the bootstrap closes on this
conjectured spectrum. Thus, unlike previously expected, we believe that
no bound states transforming under non-fundamental or reducible
representations exist.

\subsection{The connection with affine Toda solitons}\label{ssecanbpi}

As already mentioned in the introduction we conjecture that the above
constructed exact \mbox{$S$-matrices} describe the scattering of
solitons and their bound states in $\ao$ affine Toda field theories
with purely imaginary coupling constant. In this section we will
summarise briefly some of the results which support this view. 

The reason we chose trigonometric $R$-matrices as the basic building
blocks for the soliton $S$-matrices was the fact that imaginary ATFTs
are believed to display a quantum affine symmetry related to the
affine dual algebra and that the solitons transform under fundamental
representations of \uq{{\hat g}^{\vee}}. Strong  
evidence for this quantum affine symmetry stems from the work by
Bernard and LeClair as well as Felder and LeClair. In our $S$-matrix
construction we have therefore used the values of the Lorentz spins 
of the non--local conserved charges as derived in \cite{felde92}.

Possibly the strongest connection of the trigonometric $S$-matrices
with ATFTs at this stage is the breather--particle identification. By 
analogy with all previously considered cases, we expect the
$S$-matrices for the lowest breathers to coincide with the  
\mbox{$S$-matrices} for the real coupling $\ao$ ATFTs. Using formula
(\ref{anbrbrsm}) for $p=r=1$ we find
\be
S_{B_1^{(a)}B_1^{(b)}}(\th) =  \prod_{k=1}^b
\bl\frac{\la}2(b-a-2k)-1\br \bl\frac{\la}2(a+b-2k+2)\br
\bl\frac{\la}2(a+b-2k))\br \bl\frac{\la}2(b-a-2k+2)+1)\br
\label{anbr1br1} \;.
\ee
The $S$-matrix for the fundamental quantum particles in the real
theory on the other hand
was given in \cite{brade90} in the following form:
\be
S^{(r)}_{ab}(\th) = \prod_{\scs{\begin{array}{c}
			     a-b+1\\ \mbox{step } 2
			    \end{array}}}^{a+b-1}
\cbl p \cbr_r\;,
\ee
in which 
\be
\cbl y\cbr_r \equiv \frac{\bl y+1 \br_r \bl y-1 \br_r} {\bl y+1-B
\br_r \bl y-1+B \br_r}\;, \hs{1cm} \bl y\br_r \equiv
\frac{\sin(\frac{\th}{2i}+\frac{\pi y}{2(n+1)})}
{\sin(\frac{\th}{2i}-\frac{\pi y}{2(n+1)})} \;, 
\ee
and it was conjectured (and shown up to order $\beta^4$ in
perturbation theory in \cite{brade92}) that $B(\beta) =
\frac1{2\pi}\frac{\beta^2}  
{1+\beta^2/4\pi}$. If we analytically continue $\beta \to
i\beta$ we thus obtain 
\be
B(\beta) \to -\frac2{\la} \;, \hs{0.5cm} \mbox{ and} \hs{1cm} 
\bl y\br_r \to \bl\frac{\la}2 y\br\;.
\ee
Therefore, we find
\be 
S_{ab}^{(r)}(\th) \to S_{B_1^{(a)}B_1^{(b)}}(\th)\;, 
\ee
thus establishing the lowest breather--particle identification for
$a_n^{(1)}$ ATFTs. (This has also been demonstrated in
\cite{johns96}.) 

Further evidence for the $S$-matrix conjecture for affine Toda
solitons comes from the comparison of our results with some of the
results of the study of classical soliton solutions in ATFTs. 
First of all, the trigonometric $S$-matrices (without any additional
CDD factor included) display a pole structure
which is consistent with the classical soliton mass ratios.
The classical masses of the $\ao$ solitons in the $a$th multiplet are
\be 
M^{(cl)}_a = 2m \sin\(\frac{\pi a}{n+1}\)\;, \hs{2cm} (a=1,2,\dots,n)\;, 
\ee
which are the same as the quantum soliton masses
(\ref{ansolmass}). (Recall that the soliton mass ratios in theories
based on self--dual algebras are expected to remain constant when the
theory is quantised \cite{macka95}.) 

In \cite{harde95} bound states of classical affine Toda solitons were
constructed and it was found that the classical bound state spectrum
coincides with the spectrum of breathers and excited solitons
conjectured in the previous section. Note also that the conjecture of
the spectrum implies that only bound states of two solitons with equal
mass exist. This also agrees with results in classical ATFTs. 

There are of course still some serious problems regarding the
existence of a quantised version of imaginary ATFTs, which include the
fact that the Lagrangian defines an in general non--unitary theory or
that there are not enough classical solitons to fill out the
representation spaces \cite{mcghe94}. However, we believe that the
results of the construction of soliton and bound state $S$-matrices in
this and a series of other papers further support the conjecture that
there are some unitary theories embedded in the imaginary ATFTs. 
This concludes our discussion of the $\ao$ ATFTs and their
$S$-matrices. In the following section we repeat the above
construction for the case of the twisted algebra $\atn$.

%
%
\section{The $\atn$ invariant $S$-matrices}

The $\atn$ affine Lie algebras play a somewhat special role in the 
sense that they are the only non--simply--laced but nevertheless
self--dual affine Lie algebras. In this case we therefore will find
some similarities with the $d_{n+1}^{(2)}$ invariant $S$-matrices from
\cite{gande95b} as well as with those from the preceding chapter.  
We find that most of the computations used in \cite{gande95b,gande96}
can be used analogously for the case of $\atn$ and we refer the reader
to these papers for details. 

The construction of the $\atn$ $S$-matrices was made possible by the 
extension of the tensor product graph method to the case of  twisted
algebras in \cite{deliu95d}, which allowed the construction of
\uq{\atn}--invariant $R$-matrices.   
These were given in the following form: 
\be
\cR_{a,b}(x) = \sum_{c=0}^a\sum_{d=0}^c \prod_{i=c}^{a-1}\lag
a+b-2i\rag_{-} \prod_{j=1}^{c-d}\lag n-a-b+2j\rag_+
\cP_{\la_d+\la_{a+b-2c+d}}\;, \label{rmatrixatn}
\ee
in which $a,b = 1,2,\dots,n$, $a\geq b$ and a slight generalisation of
the $<,>$ bracket notation has been used: $\lag a\rag_{\pm}
\equiv \frac{1\pm xq^a}{x\pm q^a}$.

Since $\atn$ is a self--dual algebra, the connections of the
$R$-matrix parameters with the \mbox{$S$-matrix} parameters are again
given by (\ref{xqthbeta}) and (\ref{anmula}) but now with the $\atn$
Coxeter number 
\[
h = 2n+1\;.
\]

\subsection{The soliton $S$-matrices}

We make the by now familiar ansatz for a set of $\atn$ invariant soliton
$S$-matrices acting on the representation spaces of the fundamental
\uq{\atn} representations: 
\be
S_{a,b}(\theta) = F_{a,b}(\mu(\theta)) k_{a,b}(\mu(\theta))\:
\tau \cR_{a,b}(x(\th)) \tau^{-1}\;.
\label{atnsmatrix}
\ee
The fusion factor $k_{a,b}(\mu)$ is again given by (\ref{an1kab}) and
$\tau$ denotes the gauge transformation from the homogeneous to the
principal gradation. In order to find the overall scalar factor
$F_{a,b}(\mu)$ we first need the $\atn$ $R$-matrix crossing property. 
This is completely analogous to the derivation in \cite{gande95b} and
was done explicitly in the appendix of \cite{gande96c}\footnote{Note,
that some care must be taken in this case regarding the fact that
for the sake of convenience we have again chosen $q$ to differ from
(\ref{qbeta}) by a minus sign.} where it was
found that 
\be
c_{1,1}(i\pi-\th) \cR_{1,1}^{(cross)}(x(i\pi-\th)) =
c_{1,1}(\th)\cR_{1,1}(x(\th)) \;,
\ee
in which
\[
c_{1,1}(\th) = \sin(\pi(\mu-\la))\sin(\pi(\mu-(n+\frac12)\la))\;.
\]

\noi Following exactly the same method as employed in \cite{gande95b}
we find the following scalar factor in terms of Gamma functions:
\newpage
\bea
F_{1,1}(\mu) &=& \prod_{j=1}^{\infty} \frac{\Ga(\mu+jh\la-\la)
\Ga(\mu+jh\la-2n\la+1)}{\Ga(-\mu+jh\la-\la) 
\Ga(-\mu+jh\la-2n\la+1)} \nn \\
&& \hs{9pt} \times \frac{\Ga(\mu+jh\la-(n+\frac12)\la)
\Ga(\mu+jh\la-(n+\frac12)\la+1)} {\Ga(-\mu+jh\la-(n+\frac12)\la)
\Ga(-\mu+jh\la-(n+\frac12)\la+1)} \nn \\
&& \hs{9pt} \times \frac{\Ga(-\mu+jh\la-(2n+1)\la)
\Ga(-\mu+jh\la+1)}{\Ga(\mu+jh\la-(2n+1)\la) 
\Ga(\mu+jh\la+1)} \nn \\
&& \hs{9pt} \times \frac{\Ga(-\mu+jh\la-(n+\frac32)\la)
\Ga(-\mu+jh\la-(n-\frac12)\la+1)} {\Ga(\mu+jh\la-(n+\frac32)\la)
\Ga(\mu+jh\la-(n-\frac12)\la+1)}\;, \label{atnscalar}
\eea
and the higher scalar factors are again given by the fusion
formula  
\be
F_{a,b}(\mu) =\prod_{j=1}^a \prod_{k=1}^b
F_{1,1}(\mu+\frac{\la}2(2j+2k-a-b-2))\;. 
\ee
As in the previous chapter we can also rewrite this scalar factor in terms
of an integral and obtain
\be 
F_{1,1}(\mu) = \exp \bl \int_0^{\infty} \frac{dt}t 2\sinh(\mu t)
{\cal I}(t) \br \;, \label{a2nintegral}
\ee
 in which
\be
{\cal I}(t) \equiv
\frac{\sinh(\frac{\la+1}2t) \cosh((\frac{n}2-\frac14)\la t)}
{\sinh(\frac{t}2) \cosh((\frac{n}2 +\frac14)\la t)} \;.                
\ee

{}From the pole structure of the $S$-matrix we can then derive the quantum
mass ratios of the solitons. They are found to be
\be
M_a = -C 8\sqrt{2}\frac{hm}{\beta^2}\sin\( \frac{a\pi}h\)\;,
\hs{1cm} (\mbox{for } a=1,2,\dots,n)\;.
\ee
in which $C$ is some (unknown) scale factor. As expected these match
the soliton mass ratios of classical $\atn$ ATFTs.

\subsection{The bound states}

In analogy to the previous cases, the $\atn$ $S$-matrices contain
simple poles corresponding to scalar bound states, the so--called
breathers, as well as bound states transforming under fundamental
representations, the so--called excited solitons.  
Following the same procedure as before, we find the bound state poles in the
soliton $S$-matrices, as the poles at which the $S$-matrices project onto
the corresponding fundamental modules. The results of this analysis are
the following:\vs{0.2cm}
 
\noi i) The breathers $B^{(a)}_p$ are bound states of two solitons of
the same species. They correspond to the following simple poles in
$S_{a,a}(\th)$: 
\be
\mu = (n+\frac12)\la - p \;, \hs{1.5cm} (\mbox{for } a =
1,2,\dots,n\;, \mbox{ and } p=1,2,\dots\leq
(n+\frac12)\la)\;. 
\ee
{}From the position of these poles we can deduce the quantum masses of
these breathers:
\be
m_{B^{(a)}_p} = 2M_a \sin\(\frac{p\pi}{h\la}\)\;.
\ee

\noi ii) The other type of bound states correspond to the following
poles in $S_{a,a}(\th)$:   
\be
\mu = a\la -p \;, \hs{1.5cm} (\mbox{for } a =
1,2,\dots<\frac{n+1}2\;, \mbox{ and } p=1,2,\dots\leq a\la)\;, 
\ee
at which the $S$-matrix projects onto the module $V_{\la_{2a}}$. These
are the excited solitons $A_p^{(2a)}$, which only exist for even
representations. Their quantum masses are   
\be
m_{A^{(2a)}_p} = 2M_a \cos\(\frac{\pi}{h}(a-\frac{p}{\la})\)\;.
\ee
\vs{1pt}

Using these fusion poles in the bootstrap equations we can derive the
$S$-matrices for the scattering of bound states. This is again
completely analogous to the previously constructed cases and we
therefore only list the results here. We can express the $S$-matrices
in terms of a block notation:  
\be
\cbl y \cbr = \bl y\br \bl(n+\frac 12)\la -y\br \bl\la+1+y\br
\bl(n+\frac 12)\la-1-y\br\;,
\ee
in which the $\bl \br$ notation was defined in (\ref{bracket2}). Using
these notations we find the $S$-matrices for the following
scattering processes: 

\noi {\em soliton--breather scattering:}
\be
S_{A^{(a)}B^{(b)}_p}(\th)
= \prod_{k=1}^b \prod_{l=1}^p \cbl \frac{\la}2(a+b+n+\frac12
-2k)+\frac p2 -l\cbr\;,  
\ee

\noi {\em breather--breather scattering:}
\bea
S_{B_r^{(a)}B_p^{(b)}}(\th) &=& \prod_{k=1}^b \prod_{l=1}^p 
\cbl \frac{\la}2(b-a-2k)+\frac{p-r}2-l \cbr
\cbl \frac{\la}2(a+b-2k)+\frac{p+r}2-l \cbr\;,\nn \\ \label{atnBBsm}
\eea

\noi {\em excited soliton--breather scattering:}
\be
S_{A^{(2a)}_rB^{(b)}_p}(\th)
= \prod_{k=1}^b \prod_{l=1}^p \cbl \frac{\la}2(2a+b+n+\frac12
-2k)+\frac{p-r}2 -l\cbr \cbl \frac{\la}2(b+n+\frac12
-2k)+\frac{p+r}2 -l\cbr\;,  
\ee

\noi We have not attempted to construct explicit expressions of
scattering amplitudes for the scattering of two excited solitons with
each other. 

\subsection{Connection with affine Toda solitons}

In analogy to the previous chapter, we conjecture that these
$S$-matrices describe the 
scattering of quantum solitons in $\atn$ ATFTs.
All the arguments for this conjecture mentioned in section 3.6 hold
equally in this case. We also can again compare the lowest breather
$S$-matrices with the $S$-matrix for the quantum particles.
The real $\atn$ affine
Toda $S$-matrices were given in \cite{corri93} in the following form:
\be
S_{a,b}^{(r)}(\th) = \prod_{\scs{\begin{array}{c}
			     2|a-b|+2\vs{-2pt} \\ \mbox{step } 4
			    \end{array}}}^{2a+2b-2} \cbl p \cbr_r
\cbl 4n+2-p\cbr_r\;, \hs{1cm} (a,b = 1,2,\dots,n)
\ee
in which 
\be
\cbl x \cbr_r \equiv
\frac{(x-2)_r(x+2)_r}{(x-2+2B)_r(x+2-2B)_r}\;,\hs{1cm} 
\bl x\br_r \equiv \frac{\sin(\frac{\th}{2i}+\frac{\pi x}{4h})}
{\sin(\frac{\th}{2i}-\frac{\pi x}{4h})}\;. 
\ee
The coupling constant dependent function $B(\beta)$ in this case was
conjectured to be of the form $B = \frac{2\beta^2}{4\pi+\beta^2}$.
Without loss of generality let us assume $a\geq b$. Then we can
rewrite the expression for $S_{a,b}^{(r)}$ as
\[
S_{a,b}^{(r)}(\th) = \prod_{\scs{\begin{array}{c}
			      a-b+1 \vs{-2pt} \\ \mbox{step } 2
		              \end{array}}}^{a+b-1} \cbl 2p\cbr_r
\cbl 4n+2-2p\cbr_r 
= \prod_{k=1}^b\cbl 4k-2+2a-2b\cbr_r \cbl 4n+4-4k-2a+2b \cbr_r\;.
\]
If we analytically continue $\beta \to i\beta$ we find that we have to
make the following replacements:
\be
\bl y \br_r \lra \bl \frac{\la}4 y\br\;, \hs{1.5cm} B \lra -\frac2{\la}\;,
\ee
which lead to the expected result that
\be
S_{ab}^{(r)}(\th) \lra S_{B_1^{(a)}B_1^{(b)}}(\th)\;.
\ee
Thus we have established the identification of the lowest $\atn$
breathers with the $\atn$ affine Toda quantum particles.

Using the same arguments as in the previous chapter we 
conjecture that the full spectrum of $\atn$ ATFTs consists of
the elementary solitons and the above defined bound states. We expect
that all other simple and higher order poles in the $S$-matrices can
again be explained in terms of higher order diagrams using only these
states. 

Here we additionally can compare our result with another result
obtained in a different context. In 
\cite{smirn91} Smirnov discussed minimal conformal models perturbed by
the $\phi_{1,2}$ operator. This $\phi_{1,2}$--perturbed minimal model
is exactly the $a^{(2)}_2$ ATFT with imaginary coupling constant. In
the same way as we have done here, Smirnov used an
$a^{(2)}_2$--invariant 
$R$-matrix to construct the $S$-matrix for this model. Additionally,
he explicitly computes the $S$-matrix for the first breathers, which
after adjusting the different notation is indeed equal to equation
(\ref{atnBBsm}) for $n=1$. The overall $S$-matrix scalar factor in
\cite{smirn91} was given in terms of an exponential of an integral
over hyperbolic functions, which can be shown to be equal to
(\ref{a2nintegral}). This agreement with the results obtained by
Smirnov gives further support for the conjectured connection of these
$S$-matrices with the scattering of affine Toda solitons. 

In the remainder of this article we leave the subject of affine Toda
field theories and concentrate on some intriguing features of the
above constructed overall scalar factor in connection with minimal
supersymmetric $S$-matrices.

%
%
\section{Minimal supersymmetric $S$-matrices}

Exact $N=1$ and $N=2$ supersymmetric $S$-matrices in two--dimensional
integrable models have been studied by many different authors
during the last few years (see for instance
\cite{schou90,ahn90,kobay92,fendl92,lecla93b}).   
Recently some of the work on $N=1$ theories has been extended further
in \cite{moric95} and \cite{hollo97}. Apart from the usual constraints of
unitarity, crossing symmetry and the bootstrap, supersymmetric
$S$-matrices additionally are required to commute with the generators
of the supersymmetry algebra.   
One of the common features of integrable supersymmetric $S$-matrices in
two dimensions appears to be that they can be written as a 
product of a supersymmetric part and a bosonic part: 
\be
S_{a,b}(\th) = S^{(bos)}_{a,b}(\th) \ot S^{(SUSY)}_{a,b}(\th)\;,
\ee
in which the supersymmetric part seems to be universal for a large
class of different models. The entire pole structure and therefore the
spectrum of bound states is determined exclusively by the bosonic part
of the $S$-matrix. Here we only study the basic (minimal) supersymmetric
$S$-matrices without considering a specific model. In the following we
will look at two examples of sets of minimal $N=2$ and $N=1$
supersymmetric $S$-matrices.

\subsection{Minimal $N=2$ supersymmetric $S$-matrices} 

$N=2$ supersymmetric $S$-matrices were constructed by Fendley and
Intriligator in \cite{fendl92}. Without going into any detail
regarding the model they considered, we simply state their result for
the $S$-matrix.  Following the second reference in \cite{fendl92}
we consider a theory containing $2n$ different $N=2$ SUSY multiplets
denoted by $\{|u_a(\th)\rag\,,\, |d_a(\th)\rag \}$ $(a=1,2,\dots,2n)$,
which have masses 
\be
m_a = m\sin\(\frac{a\pi}{2n+1}\)\;, \label{N2mass}
\ee  
in which $m$ denotes some model specific mass parameter.
These states have rational fermion numbers ($u_a$ has fermion number
$\frac a{2n+1}$ and $d_{a}$ has fermion number $-(1-\frac a{2n+1})$)
and $u_a$ and $d_{2n+1-a}$ are charge conjugate to each other. 

Writing the incoming states in a column left of the $S$-matrix and
outgoing states in a row above, the basic form of the $S$-matrix 
appears as follows: 
\be
\begin{array}{lc}
 & \hs{-4pt} u_bu_a  \hs{24pt} d_bu_a \hs{24pt} u_bd_a \hs{24pt} d_bd_a \\ 
  \begin{array} {l}
    u_au_b \\ 
    u_ad_b \\  
    d_au_b \\  
    d_au_b 
  \end{array} & \left( \begin{array} {cccc}
                 A_{a,b}(\mu) & 0 & 0 & 0 \\
                 0 & B_{a,b}(\mu) & \tC_{a,b}(\mu) & 0 \\ 
                 0 & C_{a,b}(\mu) & \tB_{a,b}(\mu) & 0 \\
                 0 & 0 & 0 & \tA_{a,b}(\mu)
                \end{array}  \right)
\end{array} 
= S_{a,b}^{(N=2)}(\th)\;.  \label{N2sm}
\ee
This form is dictated by integrability and the requirement of fermion
number conservation. As usual, the $S$-matrix depends on the rapidity
difference of the incoming particles and we have used the abbreviation
$\mu = \frac{\th}{2i\pi}$.
The $S$-matrix elements themselves are determined (up to an overall
scalar factor) by the requirement that the $S$-matrix must commute
with the generators of the $N=2$ supersymmetry algebra. The $S$-matrix
elements can be written in the following form:  
\bea
A_{a,b}(\mu) &=&  (-)^{a+b} G_{a,b}^{(N=2)}(\mu)\;,
\nn \\
\tA_{a,b}(\mu) &=& (-)^{a+b-1}\bl - \frac{1}{2h}(a+b) \br
G_{a,b}^{(N=2)}(\mu)\;, \nn \\ 
B_{a,b}(\mu) &=& (-)^{a+b} \frac{[\frac1{2h}(b-a)]}{[\frac1{2h}(a+b)]}
G_{a,b}^{(N=2)}(\mu)\;, \nn \\
\tB_{a,b}(\mu) &=& (-)^{a+b} \frac{[\frac{1}{2h}(a-b)]}{[\frac{1}{2h}(a+b)]}
G_{a,b}^{(N=2)}(\mu)\;, \nn \\
C_{a,b}(\mu) &=& (-)^{a+b} e^{\frac{i\pi}{2h}(b-a)}
\frac{\sqrt{\sin(\frac{a\pi}{h})\sin(\frac{b\pi}{h})}}
{[\frac{1}{2h}(a+b)]} G_{a,b}^{(N=2)}(\mu)\;, \nn \\
\tC_{a,b}(\mu) &=& (-)^{a+b} e^{\frac{i\pi}{2h}(a-b)}
\frac{\sqrt{\sin(\frac{a\pi}{h})\sin(\frac{b\pi}{h})}}
{[\frac{1}{2h}(a+b)]} G_{a,b}^{(N=2)}(\mu)\;. \label{N2smel}
\eea
Here we have used the bracket notation
\be 
[y] = \sin(\pi(\mu+y)) \hs{1cm} \mbox{and} \hs{1cm} \bl y \br
=\frac{[y]}{[-y]}\;, \label{superbracket}
\ee 
which is the same as (\ref{bracket1},\ref{bracket2}) for $\la=\frac
1h$. (Note that the abbreviation $\mu = \frac{\th}{2i\pi}$ is
also equal to that used in chapter 3 for $\la=\frac 1h$.) 
The overall scalar factors $G^{(N=2)}_{a,b}(\mu)$ are determined
by $S$-matrix unitarity, crossing symmetry
and the bootstrap. They have been given in \cite{fendl92} in terms of
an infinity product of Gamma functions:
\be
G_{1,1}^{(N=2)}(\mu) = \prod_{j=1}^{\infty}\frac{\Ga(\mu+j+\frac 1h) 
\Ga(\mu+j-\frac 1h) \Ga(-\mu+j-1) \Ga(-\mu+j+1)} 
{\Ga(-\mu+j+\frac 1h) \Ga(-\mu+j-\frac 1h) \Ga(\mu+j-1) \Ga(\mu+j+1)}\;, 
\ee
and
\be
G_{a,b}^{(N=2)}(\mu) = \prod_{j=1}^a \prod_{k=1}^b
G_{1,1}^{(N=2)}(\mu+\frac 1{2(2n+1)}(a+b-2j-2k+2))\;. \label{N2scalar}
\ee
Note, however, that (\ref{N2sm}) is not yet a consistent
$S$-matrix in itself, since it contains no poles. In order to close the
bootstrap one has to include an overall CDD factor. The simplest
possible choice in this case is the minimal $a_{2n}$--Toda $S$-matrix
given by  
\be
S_{a,b}^{(min)} = \bl a+b\br \bl a+b-2\br^2 \bl a+b-4 \br^2\, \dots\,
\bl |a-b| \br\;.
\ee
This CDD factor was included in the $S$-matrix in \cite{fendl92} and
ensures that the following fusing processes are allowed: 
\[
u_a + u_b \to u_{a+b}\;, \hs{1cm} \mbox{and} \hs{1cm} d_a+d_b \to
d_{2(2n+1)-(a+b)}\;. 
\]
Without the Toda factor, the basic $S$-matrix (\ref{N2sm}) satisfies
the associated bootstrap equations passively in the sense of
\cite{hollo97}. 

In order to find a connection between these $S$-matrices and those of the
preceding chapters, we first observe that the number of particle
multiplets and the particle mass ratios (\ref{N2mass}) are the same as
those in the $a_{2n}^{(1)}$ ATFTs. 
It is known from the work in \cite{fendl91} that some
intriguing connection exists between ATFTs at one particular value 
of the coupling constant and $N=2$ supersymmetric scattering
theories.  This special value of the coupling constant is $\beta^2 =
4\pi\frac{h}{h+1}$, which in the case of $a_{2n}^{(1)}$ ATFTs is
$4\pi\frac{2n+1}{2n+2}$. In terms of the coupling constant
dependent function $\la$ this is equivalent to   
\be
\la=\frac 1h\;. \label{lah}
\ee
Furthermore, it is well known that the $N=2$ supersymmetry algebra can
be regarded as a special case of a quantum affine algebra at one
particular value of the deformation parameter $q$ (see
\cite{lecla93}). This value also corresponds to (\ref{lah}), if we
consider the relationship (\ref{xqthbeta}) of the deformation
parameter with the Toda coupling constant. 

This suggests the possibility of comparing the overall scalar factor
{}from chapter 3 (at $\la=\frac 1h$) with the $N=2$ scalar factor
(\ref{N2scalar}), and we indeed find that they are identical:
\be
G_{a,b}^{(N=2)}(\mu) = F_{a,b}(\mu)|_{\la=\frac 1h}\;,
\ee
in which $F_{a,b}(\mu)$ is given by (\ref{anFab}) with $n$ replaced by
$2n$. This demonstrates a further aspect of the relationship between
two--dimensional $N=2$ supersymmetric scattering theories and
imaginary ATFTs. 

In this paper we restrict ourselves to the discussion of this
relationship on the level of the scalar factors. It would, however, be
very interesting to examine this relationship directly on the level of
$R$-matrices. This would also require the study of $R$-matrices of
quantum affine superalgebras. The $N=2$
$S$-matrices (\ref{N2sm}), for instance, are known to be related to
the \mbox{$R$-matrices} of the superalgebra
$U_q(sl(1|1)^{(1)})$ (for further details see \cite{deliu95f}).

\subsection{Minimal $N=1$ supersymmetric $S$-matrices}

$N=1$ supersymmetric $S$-matrices for integrable theories have first been
studied in a general framework by Schoutens in \cite{schou90}. This
work has also recently been extended in \cite{moric95} and in
\cite{hollo97}. Here we consider the $S$-matrices for the scattering
of multiplets of bosons and fermions and we therefore mainly follow
the notation for the particle $S$-matrices of \cite{hollo97}. 

Let us assume we have a $N=1$ supersymmetric integrable field theory
containing $n$ multiplets of bosons and fermions, denoted by
$\{|\phi_a\rag\,,\,|\psi_a\rag\}$ ($a=1,2,\dots,n$). These particles 
have masses 
\be
m_a = m\sin(\frac{a\pi}{2n+1})\;. \label{N1mass}
\ee  
The paper \cite{hollo97} actually considered a slightly more general
case, in which $2n+1$ was replaced by an arbitrary function $H$. Since
our main purpose here is to find a connection with the above $N=2$
$S$-matrices, we will restrict ourselves to the case where $H=2n+1$. 

The minimal $N=1$ $S$-matrices, which are again determined by the
requirement that they must commute with the generators of the $N=1$
superalgebra, can be written in the following form:
\be
\begin{array}{lc}
 & \hs{-7pt} \phi_b\phi_a  \hs{23pt} \psi_b\psi_a \hs{23pt}
  \phi_b\psi_a \hs{23pt} \psi_b\phi_a \\ 
  \begin{array} {l}
    \phi_a\phi_b \\ 
    \phi_a\psi_b \\  
    \psi_a\phi_b \\  
    \psi_a\psi_b 
  \end{array} & \left( \begin{array} {cccc}
                 \cA_{a,b}(\mu) & \hs{15pt} 0 & \hs{15pt} 0 & \hs{15pt}
                 \cD_{a,b}(\mu) \\ 
                 0 & \cB_{a,b}(\mu) & \cC_{a,b}(\mu) & 0 \\ 
                 0 & \cC_{a,b}(\mu) & \tcB_{a,b}(\mu) & 0 \\
                 \cD_{a,b}(\mu) & 0 & 0 & \tcA_{a,b}(\mu)
                \end{array}  \right)
\end{array} 
= S_{a,b}^{(N=1)}(\th)\;.
\ee 
The individual scattering amplitudes in this matrix have been given as:
\bea
\cA_{a,b}(\mu) &=& \(1+\frac{2\sin(\frac{a+b}{2h}\pi)
\cos(\frac{a-b}{2h}\pi)} {\sin(\frac{\th}i)}\) G^{(N=1)}_{a,b}(\mu)\;,
\nn \\
\tcA_{a,b}(\mu) &=& \(-1+\frac{2\sin(\frac{a+b}{2h}\pi)
\cos(\frac{a-b}{2h}\pi)} {\sin(\frac{\th}i)}\) G^{(N=1)}_{a,b}(\mu)\;,
\nn \\
\cB_{a,b}(\mu) &=& \(1-\frac{2\sin(\frac{a-b}{2h}\pi)
\cos(\frac{a+b}{2h}\pi)} {\sin(\frac{\th}i)}\) G^{(N=1)}_{a,b}(\mu)\;,
\nn \\
\tcB_{a,b}(\mu) &=& \(1+\frac{2\sin(\frac{a-b}{2h}\pi)
\cos(\frac{a+b}{2h}\pi)} {\sin(\frac{\th}i)}\) G^{(N=1)}_{a,b}(\mu)\;,
\nn \\
\cC_{a,b}(\mu) &=& \frac{\sqrt{\sin(\frac{a\pi}h)
\sin(\frac{b\pi}h)}} {\sin(\frac{\th}{2i})} G^{(N=1)}_{a,b}(\mu)\;,
\nn \\ 
\cD_{a,b}(\mu) &=& \frac{\sqrt{\sin(\frac{a\pi}h)
\sin(\frac{b\pi}h)}} {\cos(\frac{\th}{2i})}
G^{(N=1)}_{a,b}(\mu)\;. \label{N1smel} 
\eea
The overall scalar factors can again be written as a product of Gamma
functions, which were given explicitly in \cite{hollo97}. For later
convenience we slightly rewrite the lowest of these factors in the
following form:  
\bea
G_{1,1}^{(N=1)}(\mu) &=&  \frac{[0][\frac 1h+\frac12]}
{[\frac 1h][-\frac 1h+\frac12]}\, \prod_{j=1}^{\infty} 
\frac{\Ga(\mu+j+\frac 1h)  
\Ga(\mu+j-\frac 1h)} {\Ga(-\mu+j+\frac 1h) \Ga(-\mu+j-\frac 1h)}
 \frac{\Ga(-\mu+j-1) \Ga(-\mu+j+1)} {\Ga(\mu+j-1) \Ga(\mu+j+1)} \nn \\
&& \hs{60pt} \times \frac{\Ga(\mu+j-\frac 12) \Ga(\mu+j+\frac 12)}
{\Ga(-\mu+j-\frac 12) \Ga(-\mu+j+\frac 12)}
\frac{\Ga(-\mu+j-\frac 1h-\frac 12) \Ga(-\mu+j+\frac 1h+\frac 12)} 
{\Ga(\mu+j-\frac 1h-\frac 12) \Ga(\mu+j+\frac 1h+\frac 12)} \;, \nn \\
\label{N1scalar}
\eea
in which we have again used the bracket notation (\ref{superbracket})
and $h=2n+1$.
All higher scalar factors can be obtained from the following fusion
formula (which can easily be derived from equation (3.17) in
\cite{hollo97}):    
\be
K_{a,b}(\mu) G_{a,b}^{(N=1)}(\mu) = \prod_{j=1}^a \prod_{k=1}^b
G_{1,1}^{(N=1)}(\mu+\frac1{2h}(a+b-2j-2k+2))\;,
\ee
in which 
\be
K_{a,b}(\mu) = \frac{[\frac 1{2h}(a+b)][\frac
1{2h}(b-a)+\frac12]}{[0][\frac12]}\, \prod_{k=1}^b 
\frac{[\frac 1{2h}(b-a-2k+2)][\frac 1{2h}(a+b-2k)+\frac12]} {[\frac
1{2h}(a+b-2k+2)][\frac 1{2h}(b-a-2k+2)+\frac 12]}\;.
\ee

By analogy with the case of the $N=2$ $S$-matrices, we first
notice that the number of multiplets and the mass ratios in the theory
are now identical to those of the $\atn$ ATFTs. We thus try to compare the
above factor $G_{a,b}^{(N=1)}$ with the $\atn$ scalar factor from
chapter 4. First we find that the factor $F_{1,1}(\mu)$ as given in
({\ref{atnscalar}) with $\la=\frac1h$ is identical to the infinite
product part in (\ref{N1scalar}). We thus have 
\be
G_{1,1}^{(N=1)}(\mu) = \frac{[0][\frac 1h+\frac12]}
{[\frac 1h][-\frac 1h+\frac12]}
F_{1,1}(\mu)|_{\la=\frac 1h}\;.
\ee
Using the above fusion formula we can easily extend this to the case
of general $a$ and $b$ and obtain
\be
G_{a,b}^{(N=1)}(\mu) = \Lambda_{a,b}(\mu) F_{a,b}(\mu)\;,
\ee
in which
\be 
\Lambda_{a,b}(\mu) = (-)^b \frac{[0][\frac 12]}{[\frac
1{2h}(a+b)][\frac 1{2h}(b-a)+\frac12]}\, \prod_{k=1}^b \( \frac
1{2h}(a-b+2k)+\frac 12\)\;.
\ee
Thus, we have established a connection between the $S$-matrices
constructed from \uq{\atn} invariant $R$-matrices and the minimal $N=1$
supersymmetric particle $S$-matrices. 

As mentioned before, it would be interesting to explore this
relationship further and examine the underlying connection between the
$N=1$ superalgebra and the quantum algebra \uq{\atn}.
It would be interesting to explore whether similar connections can be
obtained for the trigonometric $S$-matrices based on other non--simply
laced or twisted algebras. We hope to address these issues in a future
publication. In the following section we will proceed by establishing a
relationship between the $N=2$ and the $N=1$ $S$-matrix scalar factors
among each other.  

\subsection{Folding from $N=2$ to $N=1$}

In \cite{melze94} Melzer conjectured a connection between the $N=2$
and $N=1$ scattering theories via a ``folding'' of their TBA
systems (see also \cite{andre97}). In \cite{moric95}
Moriconi and Schoutens completed the Thermodynamic Bethe Ansatz for
these $N=1$ supersymmetric theories and were able to prove Melzer's
conjecture. This folding procedure takes the following form: \vs{3pt}
\\
If $\Phi_{a,b}$ denotes
the kernel of the TBA system of the $N=2$ scattering theories
discussed in section 5.1, then one can define a folded kernel by
\be 
\Phi^{folded}_{a,b} = \Phi_{a,b} + \Phi_{a,2n+1-b}\;, \hs{1cm}
(\mbox{for} \hs{0.3cm} a,b = 1,2,\dots,n)\;. \label{tbafold}
\ee
It turns out that $\Phi^{folded}_{a,b}$ is exactly the kernel of the
TBA system of the $N=1$ theories as discussed in section 5.2. 
This immediately raises the question, whether a similar folding
procedure can be established directly on the level of 
$S$-matrices. As a small step towards the answer of this question,
here we will show that the overall scalar factors satisfy a similar
folding identity.  

First note that the kernel of a TBA system is related to the logarithm
of the $S$-matrix and therefore the sum in (\ref{tbafold}) should
become a product in terms of $S$-matrices. We then use the explicit
expressions for the overall scalar factors as given above and obtain
the following identity 
\be
G_{1,1}^{(N=2)}(\mu)G_{1,2n}^{(N=2)}(\mu) = - \frac{[\frac 1h]}{[0]}
G_{1,1}^{(N=1)}(\mu)\;, 
\ee
Using the fusion relations and the connections with the affine Toda
scalar factors at $\la=\frac 1h$ we obtain after a somewhat tedious
but straightforward computation 
\be
G_{a,b}^{(N=2)}(\mu)G_{a,2n+1-b}^{(N=2)}(\mu) =
- \frac{[\frac1{2h}(a+b)][\frac1{2h}(b-a)+\frac12]} {[0][\frac12]}\,
G_{a,b}^{(N=1)}(\mu)\;. 
\ee
These relations between the $N=2$ and $N=1$ scalar factors provide the
first step towards establishing Melzer's folding from $N=2$ to
$N=1$ theories directly on the level of $S$-matrices. 
Although the above folding relation for the scalar
factors is encouraging, it is still very hard to see how a similar
folding relation could be possible for the $S$-matrices.
Such a relation could obviously not be just a product of two
$S$-matrices due to the non--diagonality of the matrices. 
It is also hard to see how the states in the theories could be
related, given the fact that the $N=2$ theory in \cite{fendl92} was
formulated in terms of fractionally charged states, whereas the $N=1$
$S$-matrices describes the scattering of bosons and fermions. There are
certainly no obvious connections between the $S$-matrix elements
(\ref{N2smel}) and (\ref{N1smel}).

As shown in the preceding sections the overall scalar factors of the
supersymmetric $S$-matrices are related to those of the $a_{2n}^{(1)}$
and $\atn$ invariant $S$-matrices. These two algebras have the same
Coxeter number $h=2n+1$ and the solitons in the
related ATFTs have the same mass ratios. The main difference between
the two theories is the fact that the $a_{2n}^{(1)}$ theory contains
$2n$ particle multiplets which occur in mass--degenerate pairs ($m_a =
m_{2n+1-a}$), whereas the $\atn$ theory contains only $n$
(non--degenerate) multiplets. Since each multiplet in ATFT can be
associated with a spot in the corresponding Dynkin diagram, it seems
plausible that any $S$-matrix folding should somehow be related to the
folding of Dynkin diagrams. We therefore expect the folding from
$N=2$ to $N=1$ theories to be related to the folding of the Dynkin
diagram of $a_{2n}^{(1)}$ to that of $\atn$. This folding of the
Dynkin diagrams was described in \cite{khast95}.  However,
a more detailed discussion of this will have to wait for a
future publication.

\vs{1.5cm} 

\noi{\bf \ul{Acknowledgements:}}\\
I would like to thank Niall MacKay, Gerard Watts, Avijit Mukherjee and
Marc Grisaru for discussions. I would also like to thank the Brandeis
Physics Department and Marc Grisaru for their hospitality while part
of this work was carried out.  This work was partly funded by a
research fellowship within the NATO-program of the German Academic
Exchange Service (DAAD).

\renewcommand{\baselinestretch}{1.0}

\end{document}